\documentclass[12pt,onecolumn,oneside,a4paper,peerreview]{IEEEtran_kit}

\usepackage{cite}
\usepackage{float}
\usepackage{graphicx}
\usepackage{amsmath}
\usepackage{times}
\usepackage{graphicx}
\usepackage{latexsym}
\usepackage{graphicx}
\usepackage{bm}
\usepackage{amssymb}
\usepackage[center]{caption2}
\usepackage{stfloats}
\usepackage{cases}
\usepackage{array}
\usepackage{setspace}
\usepackage{fancyhdr}
\usepackage{color}
\usepackage{subfigure}
\usepackage{citesort}
\usepackage{multirow}
\usepackage{amsmath}
\usepackage{cases}
\usepackage[noend]{algpseudocode}
\usepackage{algorithm}
\usepackage{algorithmicx}

\usepackage{diagbox}
\usepackage{algpseudocode}
\usepackage{epstopdf}
\usepackage{varwidth}
\usepackage{pifont}

\def\bm#1{\mbox{\boldmath $#1$}}

\renewcommand{\baselinestretch}{0.97}

\begin{document}
\title{An Attention-Aided Deep Learning Framework for Massive MIMO Channel Estimation}
 \author{\authorblockN{Jiabao Gao, Mu Hu, Caijun Zhong, Geoffrey Ye Li, and Zhaoyang Zhang
 \thanks{J. Gao, M. Hu, C. Zhong, Z. Zhang are with the College of Information Science and Electronic Engineering, Zhejiang University, Hangzhou, China (Email:  \{gao\_jiabao, muhu, caijunzhong\}@zju.edu.cn). Geoffrey Ye Li is with the Faculty of Engineering, Department of Electrical and Electronic Engineering, Imperial College London, England (Email: Geoffrey.Li@imperial.ac.uk).}
 }}
\maketitle
\begin{abstract}
Channel estimation is one of the key issues in practical massive multiple-input multiple-output (MIMO) systems. Compared with conventional estimation algorithms, deep learning (DL) based ones have exhibited great potential in terms of performance and complexity. In this paper, an attention mechanism, exploiting the channel distribution characteristics, is proposed to improve the estimation accuracy of highly separable channels with narrow angular spread by realizing the ``divide-and-conquer" policy. Specifically, we introduce a novel attention-aided DL channel estimation framework for conventional massive MIMO systems and devise an embedding method to effectively integrate the attention mechanism into the fully connected neural network for the hybrid analog-digital (HAD) architecture. Simulation results show that in both scenarios, the channel estimation performance is significantly improved with the aid of attention at the cost of small complexity overhead. Furthermore, strong robustness under different system and channel parameters can be achieved by the proposed approach, which further strengthens its practical value. We also investigate the distributions of learned attention maps to reveal the role of attention, which endows the proposed approach with a certain degree of interpretability.
\vspace{1cm}
\begin{center}
{\bf Index Terms}
\end{center}
Massive MIMO, channel estimation, deep learning, attention mechanism, hybrid analog-digital, divide-and-conquer.
\end{abstract}

\newpage

\section{Introduction}

Massive multiple-input multiple-output (MIMO) is a key enabling technology for future wireless communication systems due to its high spectral and energy efficiency\cite{massiveMIMO1,massiveMIMO2}. However, the realization of various theoretical gains of massive MIMO is critically dependent on the quality of channel state information (CSI). Because of the large number of antennas and users, the CSI acquisition has long been a major challenge in practical massive MIMO systems.

In the prior works, least square (LS) and minimal mean-squared error (MMSE)\cite{LS_MMSE} are two most commonly used estimators for channel estimation. The LS is relatively simple and easy to implement while its performance is unsatisfactory. On the other hand, MMSE can refine the LS estimation if accurate channel correlation matrix (CCM) is available. However, the complexity of MMSE estimation is much higher than that of LS estimation due to the matrix inversion operation. On the other hand, to reduce the hardware and energy cost, the hybrid analog-digital (HAD) architecture is usually adopted in practical massive MIMO systems, where the multi-antenna array is connected to only a limited number of radio-frequency (RF) chains through phase shifters in analog domain\cite{HAD1,HAD2,new}. With HAD, channel estimation becomes even more difficult since the received signals at the BS are only a few linear combinations of the original signals. If LS is used, multiple estimations are required since only part of the antennas' channels can be estimated once due to limited number of RF chains. To avoid the dramatically increased overhead of LS, the slowly changing directions of arrival of channel paths are obtained first in the preamble stage in \cite{HAD_angular}, then only channel gains of each path are re-estimated in a long period. Another alternative is to exploit the channel sparsity and estimate all antennas' channels at once using the compressed sensing (CS) based methods, such as orthogonal matching pursuit\cite{HAD_CS1} and sparse Bayesian learning\cite{HAD_CS2}. In \cite{Structured_SBL,VBI}, several improved CS algorithms have been developed through embedding the structural characteristics of channel sparsity, which can achieve better estimation performance without extra pilot overhead. Nevertheless, CS algorithms require high computational complexity and perform poor for channels with low sparsity. Therefore, it is highly desirable to develop channel estimators with less requirement for prior information and better performance-complexity trade-off.

Inspired by the great performance and the low complexity during online prediction, deep learning (DL) has been applied to many wireless communication problems\cite{survey,CGAN}, such as spectrum sensing\cite{SpectrumSensing}, resource management\cite{ResourceManagement1,NewRef_Res1,NewRef_Res2,NewRef_Res3}, beamforming\cite{Beamforming1,Beamforming2}, signal detection\cite{Powerof,model_driven,NewRef_SD}, and channel estimation\cite{DL_CE1,LDAMP,LISTA,DL_CE2,DL_CE3,DL_CE4,modal,DLCS,DL_CE5}. By exploiting the structural characteristics of the modulated signals, the customized deep neural network (DNN) in \cite{SpectrumSensing} significantly outperforms energy detection in spectrum sensing. In \cite{ResourceManagement1}, a DNN has been proposed for resource management, which can achieve comparable performance as the iterative optimization algorithm. An unsupervised learning-based beamforming network has been developed for intelligent reconfigurable surface aided massive MIMO systems in \cite{Beamforming1}. In \cite{Powerof}, channel estimation and signal detection in orthogonal frequency division multiplexing systems have been performed jointly by a DNN. Then, a model-driven based approach is further proposed in \cite{model_driven} to exploit the advantages of both conventional algorithms and DNN. In \cite{NewRef_SD}, rather than directly using a black-box DNN, the conventional orthogonal approximate message passing algorithm (OAMP) is unfolded for the detection network.

There are mainly two categories of approaches for DL-based massive MIMO channel estimation. In the first category, ``deep unfolding" methods unfold various iterative optimization algorithms and enhance their estimation performance by inserting learnable parameters. In \cite{LDAMP}, the AMP algorithm is unfolded into a cascaded neural network for millimeter wave channel estimation, where the denoiser is learned by a DNN. Thanks to the power of DL, the proposed method can outperform a series of conventional denoising-AMP based algorithms. In \cite{LISTA}, the iterative shrinkage thresholding algorithm is unfolded to solve sparse linear inverse problems, where massive MIMO channel estimation is used as a case study. However, ``unfolding" is only feasible to the iterative algorithms with simple structures, and the computational complexity is also high. In the other category, DL is used to directly learn the mapping from available channel-related information to the CSI for performance improvement or complexity reduction. In \cite{DL_CE1}, a DNN has been proposed to refine the coarse estimation in HAD massive MIMO systems, where the channel correlation in the frequency and time domains is exploited for further performance improvement. In \cite{DL_CE3}, the estimation performance is further improved by jointly training the pilot signals and channel estimator with an autoencoder in downlink massive MIMO systems. In \cite{DL_CE4}, graph neural network has been used for massive MIMO channel tracking. Deep multimodal learning has been used for massive MIMO channel estimation and prediction in \cite{modal}. To reduce the complexity, the amplitudes of beamspace channels are predicted by a DNN and the dominant entries are estimated by LS in \cite{DLCS}, thus avoiding the greedy search commonly adopted by CS algorithms. In \cite{DL_CE5}, the uplink-to-downlink channel mapping in frequency-division duplex (FDD) systems is learned by a sparse complex valued network. 

Nevertheless, current DL-based channel estimation methods have seldom exploited the characteristics of channel distribution. In practice, the BS is often located in a high altitude with few surrounding scatters\cite{narrow_AS}, so the angular spread of each user's incident signal at the BS is narrow. Thus, the global distribution of channels corresponding to different users in the entire angular space can be viewed as the composition of many local distributions, where each local distribution represents channels within a small angular region. Due to narrow angular spread, a certain angular region contains much fewer channel cases than the entire angular space because of the limited angular range of channel paths, making the local distributions much simpler than the global distribution. Besides, different local distributions can be highly distinguishable from each other if the entire angular space is properly segmented into different angular regions. Under such a condition, the classic ``divide-and-conquer" policy, which tackles a complex main problem by solving a series of its simplified sub-problems, is very suitable. Specifically, the estimation of channels in the entire angular space can be regarded as the main problem and the estimation of channels in different small angular regions can be regarded as different sub-problems. Motivated by this, in this paper, we propose a novel attention-aided DL-based channel estimation framework, where the ``divide-and-conquer" policy is realized automatically through the dynamic adaptation of attention maps. The main contributions of this paper are summarized as follows:
\begin{itemize}
\item An attention-aided DL-based channel estimation framework is proposed for massive MIMO systems, which achieves better performance than its counterpart without attention in simulation. To the best knowledge of the authors, this is the first work that introduces the attention mechanism to DL-based channel estimation\footnote{{There are already some literature that uses attention-aided DL to solve communication problems, such as CSI compression\cite{attention_compression1,attention_compression2} and joint source and channel coding\cite{attention_SNR}. Nevertheless, the considered channel distribution in \cite{attention_compression1} does not possess strong separable property, and the proposed method in \cite{attention_SNR} requires extra side information. As for \cite{attention_compression2}, the non-local neural network model is utilized to exploit the self-attention in the spatial dimension of channels.}}.

\item We extend the above framework to the scenario with HAD and an embedding method is proposed to effectively integrate the attention mechanism into the fully connected neural network (FNN), which expands the application range of the proposed approach.

\item We visually explain the ``divide-and-conquer" policy reflected in the distributions of learned attention maps, which enhances the interpretability and rationality of the proposed approach.

\item Based on our results, the performance gain of attention mainly comes from the narrow angular spread characteristic of channels. Therefore, the proposed approach can be extended to many other problems apart from channel estimation as long as the channel distribution has certain separability, such as multi-user beamforming, FDD downlink channel prediction, and so forth.
\end{itemize}

The rest of this paper is organized as follows. Section II introduces the system model, channel model, and problem formulation. Section III presents the attention-aided DL-based channel estimation framework, which is extended to the HAD scenario in Section IV. Extensive simulation results are demonstrated in Section V. Eventually, the paper is concluded in Section VI.

Here are some notations used subsequently. We use italic, bold-face lower-case and bold-face upper-case letter to denote scalar, vector, and matrix, respectively. ${\bf A}^{T}$ and ${\bf A}^{H}$ denote the transpose and Hermitian or complex conjugate transpose of matrix $\bm{A}$, respectively. $[\bm{A}]_{i,j}$ denotes the element at the $i$-th row and $j$-th column of matrix $\bm{A}$. ${\left\| {\bf{x}} \right\|}$ denotes the $l$-2 norm of vector $\bm{x}$, and $|a|$ denotes the amplitude of complex number $a$. ${\mathbb C^{x \times y}}$ denotes the ${x \times y}$ complex space. $\mathcal{CN}(\mu,\sigma^2)$ denotes the distribution of a circularly symmetric complex Gaussian random variable with mean $\mu$ and variance $\sigma^2$. $\mathcal{U}[a,b]$ denotes the uniform distribution between $a$ and $b$.

\section{System model and problem formulation}
In this section, system model and channel model are first introduced. Then, the conventional massive MIMO channel estimation problem is formulated.
\subsection{System Model}
\begin{figure}[htbp]
\centering
\includegraphics[width=1\textwidth]{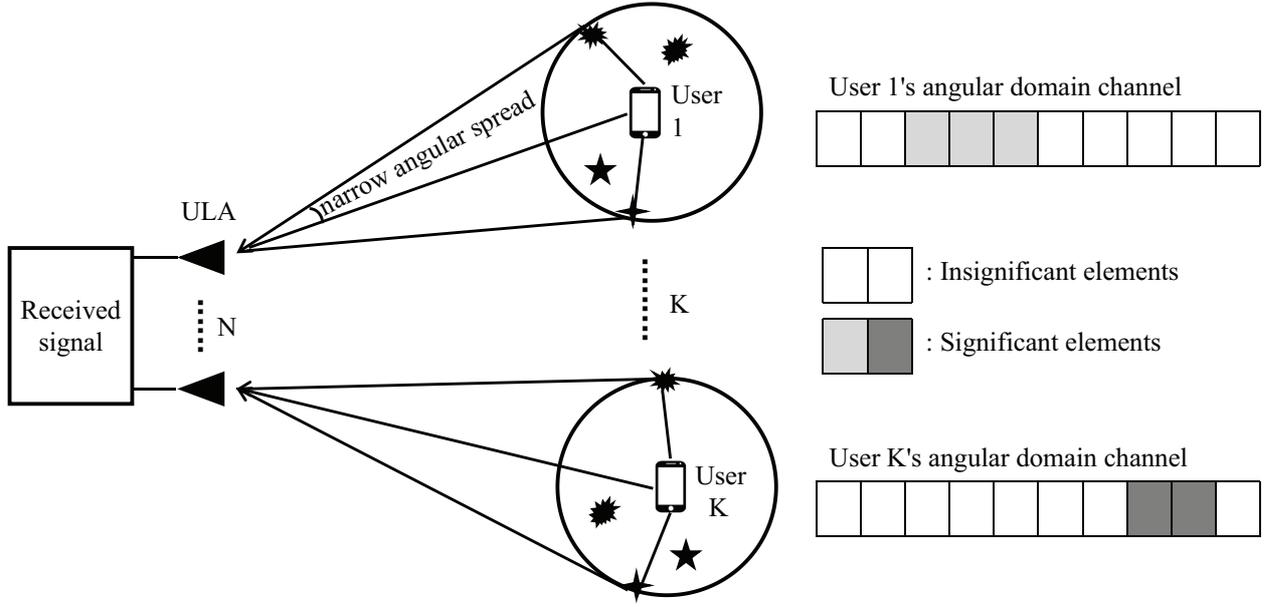}
\caption{Massive MIMO system without HAD.}
\label{system}
\end{figure}

Consider a single cell massive MIMO system, where the BS is equipped with an $N$-antenna uniform linear array (ULA) and $K$ single-antenna users are randomly distributed in the cell of the corresponding BS, as illustrated in Fig. \ref{system}.

\subsection{Channel Model}
Following the same channel model as in \cite{Unified}, the uplink channel from user $k$ to the BS can be expressed as
\begin{equation}
\bm{h}_k=\frac{1}{\sqrt{N_p}}\sum_{i=1}^{N_p}\alpha_{ki}\bm{a}(\theta_{ki})\in \mathbb{C}^{N \times 1},
\label{channel}
\end{equation}
where $N_p$ is the number of paths, $\alpha_{ki}$ and $\theta_{ki}$ are the complex gain and angle of arrival (AoA) at the BS of the $i$-th path from the $k$-th user, respectively. Without loss of generality, we consider half-wavelength antenna spacing in this paper, then the steering vector of the ULA can be written as $\bm{a}(\theta)=[1,e^{j\pi\sin(\theta)},\cdots,e^{j\pi\sin(\theta)(N-1)}]^T$. Define the average AoA and the angular spread of user $k$'s channel paths as $\bar{\theta}_k$ and $\bigtriangleup_{\theta}$, respectively, that is, $\theta_{ki}$ follows a uniform distribution $\mathcal{U}[\bar{\theta}_k-\bigtriangleup_{\theta},\bar{\theta}_k+\bigtriangleup_{\theta}]$. As in \cite{Unified,VBI}, the narrow angular spread assumption is adopted, i.e., $\bigtriangleup_{\theta} \ll \pi$. 

To better understand this channel characteristic, we convert the original channel to the angular domain by
\begin{align}
\bm{x}_k=\bm{F}\bm{h}_k\in \mathbb{C}^{N \times 1},
\end{align}
where $\bm{x}_k$ denotes the angular domain channel of user $k$, and $\bm{F}\in\mathbb{C}^{N \times N}$ is a shift-version discrete Fourier transform matrix\cite{VBI}, with the $n$-th row given by $\bm{f}_{n}=\frac{1}{\sqrt{N}}[1,e^{-j\pi \eta_n},\cdots,e^{-j\pi \eta_n(N-1)}]$, for $\eta_n = \frac{-N+1}{N},\frac{-N+3}{N},\cdots,\frac{N-1}{N}$. Due to narrow angular spread assumption, the angular domain channel exhibits the spatial-clustered sparsity structure\cite{VBI}. Specifically, as shown in the right half of Fig. \ref{system}, $\bm{x}_k$ only has a few significant elements appearing in a cluster. If properly exploited, such sparsity structure can help to improve estimation performance and reduce estimation overhead.

\subsection{Problem Formulation}
During the uplink training, orthogonal pilot sequences are sent by different users. Denote the pilot sequence of the $k$-th user as $\bm{p}_k\in{\mathbb C}^{1\times Lp}$, where $L_p\ge K$ is the length of pilot sequences. Notice that the channel during pilot training phase is assumed to be unchanged\cite{VBI} since $L_p$ is relatively small. Therefore, the superimposed received signal at the BS can be expressed as
\begin{align}
\bm{Y}=\sum^K_{k=1}\bm{h}_k\bm{p}_k+\bm{N}\in \mathbb{C}^{N \times L_p},
\end{align}
where $\bm{N}\sim \mathcal{CN}(0,\sigma^2)\in{\mathbb C}^{N\times L_p}$ is the zero-mean additive white Gaussian noise at the BS with variance $\sigma^2$.  Without loss of generality, we fix the power of pilot sequences to unit and adjust the transmit signal-to-noise ratio (SNR) by changing the noise variance. Then, we have $\bm{p}_i\bm{p}_j^H=0, \forall i \neq j$ and $\bm{p}_i\bm{p}_i^H=1, \forall i$. Exploiting the orthogonality of the pilot sequences, the LS estimation of user $k$'s channel can be obtained as
\begin{equation}
\hat{\bm{h}}_k = \bm{Y}\bm{p}_k^H = \bm{h}_k+\widetilde{\bm{n}}_{k}\in \mathbb{C}^{N \times 1},
\label{receive}
\end{equation}
where $\widetilde{\bm{n}}_{k}\triangleq \bm{N}\bm{p}_k^H$ is the effective noise for user $k$. For brevity, we will consider a specific user from now on and omit subscript $k$. Besides, we use $\hat{\bm{h}}_\text{LS}$ to denote the LS estimation. Therefore, the goal of channel estimation\footnote{Here we use the term ``channel estimation" for consistency, actually ``channel refinement" is more proper.} is to find a function that maps $\hat{\bm{h}}_\text{LS}$ to $\bm{h}$.

One of the conventional methods is the MMSE estimation, where the LS estimation is refined by the CCM. However, accurate CCM is hard to obtain in practice and the complexity of matrix inversion in MMSE estimation is very high, especially when the antenna number is large. In \cite{YYW}, DL-based methods have been proposed to refine the channel estimation. In this paper, we will develop an attention-aided DL framework for conventional massive MIMO channel estimation by exploiting the characteristics of channel distribution.

\section{Attention-aided DL framework for massive MIMO channel estimation}
In this section, input and output processing, network structure design, and detailed network training method of the proposed framework are introduced.

\subsection{Input and Output Processing}
Since channel parameters can be canonically expressed in the angular domain, the input and output of the networks are all in the angular domain in the proposed framework. In simulation, we find that the more sparse angular domain input and output can lead to better channel estimation performance than the original ones. Once the angular domain channel estimation, $\hat{\bm{x}}$, is obtained, the original channel estimation can be readily recovered by $\hat{\bm{h}}=\bm{F}^H\hat{\bm{x}}$. Besides, the real and imaginary parts have to be separately processed since complex training is still not well supported by current DL libraries. To promote efficient training, we also perform standard normalization preprocessing on the input. 

\subsection{Attention-Aided Channel Estimation Network Structure Design}
\begin{figure}[htbp]
\centering
\includegraphics[width=1\textwidth]{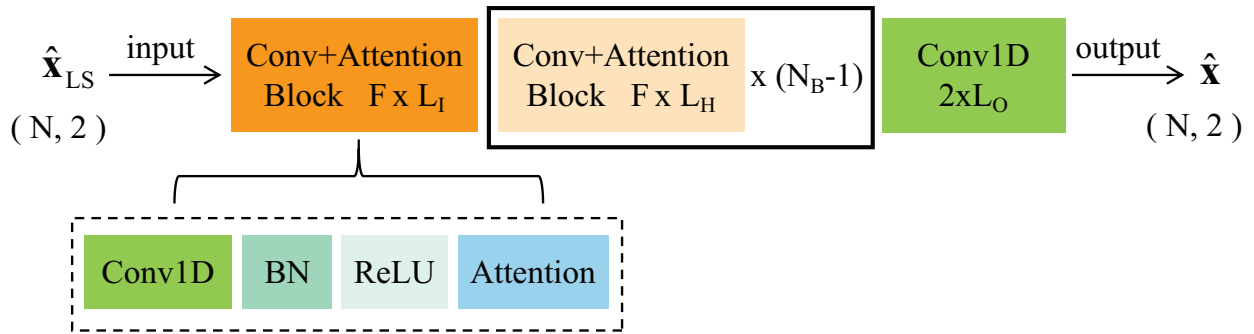}
\caption{Structure of the channel estimation network.}
\label{net1}
\end{figure}

As shown in Fig. \ref{net1}, convolutional neural network (CNN) is a suitable choice for the network structure to exploit the local correlation in the input data due to the spatial-clustered sparsity structure of the angular domain channel. In this paper, one-dimensional convolution (Conv1D) is used due to the shape of input data. The input of a Conv1D layer is organized as a $(F,C)$-dimensional feature matrix, where $C$ denotes the number of channels\footnote{Here channel is a term in CNN representing a dimension of feature matrix, not the communication channel.} and $F$ denotes the number of features in each channel. Then, the convolution operation slides $C^\prime$ filters over the input feature matrix in certain strides to obtain the output feature matrix, which is also the input of the next layer. Specifically, each filter contains a $(L,C)$-dimensional trainable weight matrix and a scalar bias term, where $L$ denotes the filter size. When a filter is located in a certain position of the feature matrix, the cross-correlation between the corresponding chunk of the feature matrix and the weight matrix of the filter is computed and the bias is added to obtain the convolution output of the position\cite{DL_concepts}. In the proposed channel estimation network, $N_B$ convolutional blocks and an output Conv1D layer are used to refine the LS coarse channel estimation. As depicted in the dashed box, in each convolutional block, a batch normalization (BN) layer to prevent gradient explosion or vanishing\cite{BN} and a ReLU activation function are inserted after the Conv1D layer. Besides, the Conv1D layer in the first block has $F$ filters of size $L_I$ and the Conv1D layers in the next $N_B-1$ blocks have $F$ filters of size $L_H$. The optimal values of $N_B$ and $F$ can be determined through simulation. Finally, the output Conv1D layer has 2 filters of size $L_O$, corresponding to the real and imaginary parts of the channel prediction, respectively. The stride is set to $S$ and all the Conv1D layers pad zeros to keep the dimension $N$ of the feature matrix unchanged.

To effectively exploit the distribution characteristics of channel, the attention mechanism\footnote{Notice that, the term attention can refer to many related methods including \cite{attention,dual_attention,non_local}. In this paper, we use the classic ``SENet" proposed in \cite{attention}.} is applied in the network structure design. In the original CNN, all the features are used for all data samples with equal importance. However, certain features can definitely be more important or informative than others to certain data samples in practice, especially for highly separable data like narrow angular spread channel. For instance, key features, which are only aimed at dealing with channel distribution in a specific angular region, might be useless or even disruptive for the estimation of channels in another region far apart. Therefore, the idea of feature importance reweighting can be used here to improve network performance.

\begin{figure}[htbp]
\centering
\includegraphics[width=1\textwidth]{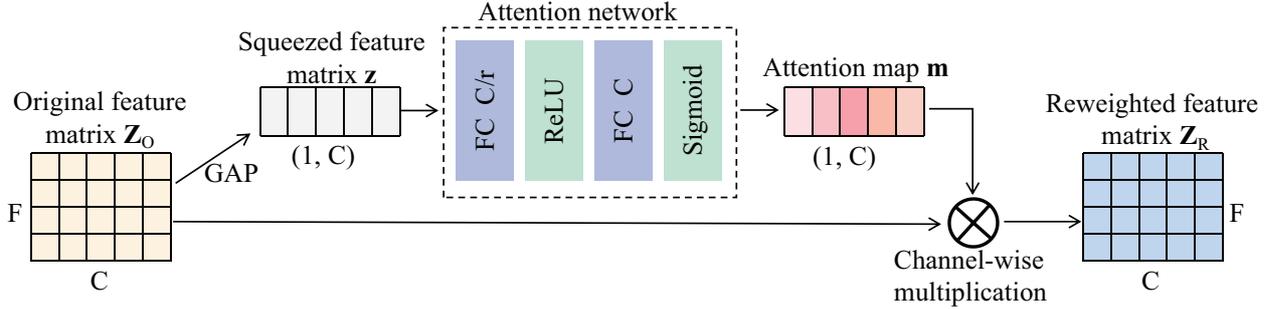}
\caption{Structure of the attention module.}
\label{attention}
\end{figure}

As is demonstrated in Fig. \ref{attention}, the original feature matrix is multiplied by an attention map in a channel-wise manner to obtain the reweighted feature matrix in the attention module, where more important or informative features to the current data sample will be paid more ``attention" to. For the learning process of the attention map, global average pooling is performed first on the original feature matrix, $\bm{Z}_O$, to embed the global information into a $(1,C)$-dimensional squeezed feature matrix, $\bm{z}$. Specifically, the $c$-th element of $\bm{z}$ is calculated by $z_c=\sum_{f=1}^F[\bm{Z}_O]_{f,c}/F$\cite{attention}. Then, the $(1, C)$-dimensional attention map, $\bm{m}$, is predicted by a dedicated attention network based on $\bm{z}$. The attention network contains two fully connected (FC) layers. The first FC layer with $C/r$ neurons is followed by a ReLU activation, $f_\text{ReLU}(x)=\text{max}(0,x)$, where $r\ge 1$ denotes the reduction ratio. The second FC layer with $C$ neurons is followed by a Sigmoid activation, $f_\text{Sigmoid}(x)=1/(1+e^{-x})$, which limits the elements of $\bm{m}$ between 0 and 1. As can be seen in Fig. \ref{net1}, an attention module is inserted at the end of each convolutional block in the proposed channel estimation network. Besides, $r$ is set to 2 to balance performance and complexity and the FC layers in the attention network do not use bias to facilitate channel dependency modeling.

\subsection{Network Training}
To train the designed network, the mean-squared error (MSE) between the true angular domain channel, $\bm{x}$, and the predicted angular domain channel, $\hat{\bm{x}}$, is used as the loss function, which can be calculated by
\begin{equation}
\text{MSE Loss}=\frac{1}{n}\sum_{i=1}^{n}\left\|\hat{\bm{x}}_i-\bm{x}_i\right\|^2,
\end{equation}
where subscript $i$ denotes the $i$-th data sample in a mini-batch and $n=500$ is the size of the mini-batch. Xavier\cite{xavier} is used as the weight initializer and Adam\cite{adam} is used as the weight optimizer. The initial learning rate is set to 0.001. To balance the training complexity and testing performance, we generate totally 200,000 data samples according to the adopted channel and transmission models. Then, the generated dataset is split into training, validation, and testing set with a ratio of 3:1:1. In order to accelerate loss convergence at the beginning and reduce loss oscillation near the end of training, the learning rate is set to decay 10 times if the validation loss does not decrease in 10 consecutive epochs. Besides, early stopping \cite{early_stopping} with a patience of 25 epochs is applied to prevent overfitting and speed up the training process.

\section{Extension to the HAD scenario}
\begin{figure}[htbp]
\centering
\includegraphics[width=0.8\textwidth]{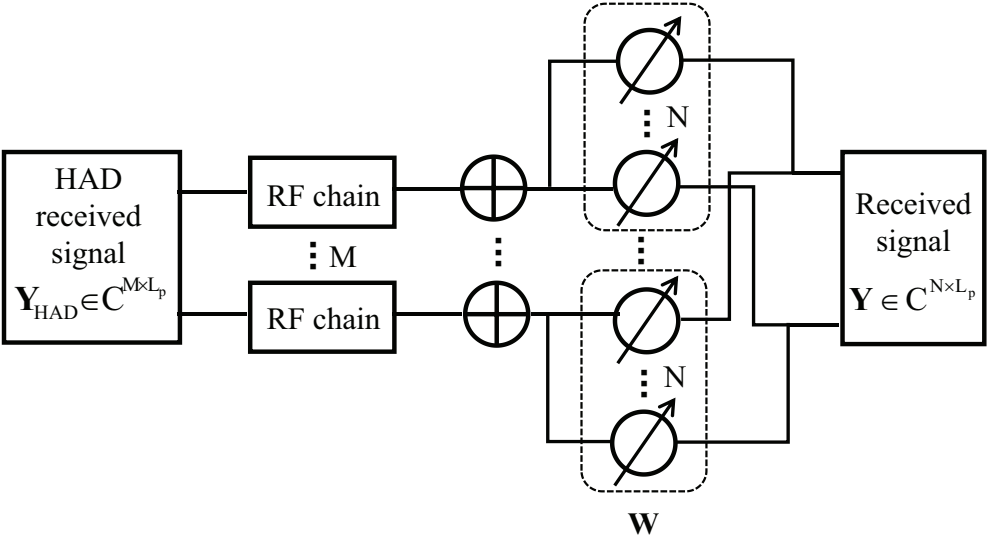}
\caption{Massive MIMO system with HAD.}
\label{system_HAD}
\end{figure}

In practice, the HAD architecture is often adopted in massive MIMO systems to save hardware and energy cost. Due to the effect of phase shifters in the analog domain in the HAD architecture, the problem formulation of channel estimation changes and the channel estimation network structure has to be customized correspondingly as well. In the HAD architecture, we assume there is only $M \ll N$ RF chains available at the BS, as illustrated in Fig. \ref{system_HAD}.

\subsection{Problem Reformulation with HAD}
With HAD, the signals arriving at the antennas have to go through the phase shifters first before received by the RF chains. So, the eventual received signal on the baseband can be expressed as
\begin{align}
\bm{Y}_\text{HAD}=\bm{WY}\in \mathbb{C}^{M \times L_p},
\end{align}
where $\bm{W}\in{\mathbb C}^{M\times N}$ denotes the analog combining matrix. As the phase shifters only change the phase of signals, we have $|[\bm{W}]_{i,j}|=1/\sqrt{N}$, $\forall i,j$ after normalization. We set $\bm{W}$ to a matrix whose rows are length-$N$ Zadoff-Chu sequences with different shifting steps as in\cite{Structured_SBL}. Again, exploiting the orthogonality of the pilot sequences, the received signal corresponding to user $k$ can be obtained as
\begin{equation}
\bm{y}_k = \bm{Y}_\text{HAD}\bm{p}_k^H = \bm{Wh}_k+\widetilde{\bm{n}}^\prime_{k}\in \mathbb{C}^{M \times 1},
\label{receive_HAD}
\end{equation}
where $\widetilde{\bm{n}}^\prime_{k}\triangleq \bm{W}\widetilde{\bm{n}}_{k}$ is the effective noise for user $k$ with HAD. Consider a specific user and omit the subscript $k$, the goal of channel estimation now becomes to find a function that maps $\bm{y}$ to $\bm{h}$.

Since the overhead of LS estimation increases dramatically due to limited number of RF chains, CS algorithms are more often adopted to solve the channel estimation problem in HAD massive MIMO systems conventionally. However, the performance of CS algorithms is highly dependent on channel sparsity and the computational complexity is relatively high due to complex operations and a large number of iterations. Therefore, we extend the proposed framework to the HAD scenario and use DL to overcome these issues.

\subsection{Attention-Aided Channel Estimation Network Structure Design With HAD}
Different from the former scenario, in the problem of channel estimation with HAD, the input data becomes the received signal $\bm{y}$, where little local correlation exists due to the compression of matrix $\bm{W}$. Therefore, FNN should be used rather than CNN to achieve better performance. Although the attention mechanism has been originally proposed in the area of computer vision and is only compatible with CNN, its key idea, feature importance reweighting, is actually independent of network structure. Therefore, to exploit the benefit of the attention mechanism, we propose a simple but effective method to embed it into FNN.

\begin{figure}[htbp]
\centering
\includegraphics[width=1\textwidth]{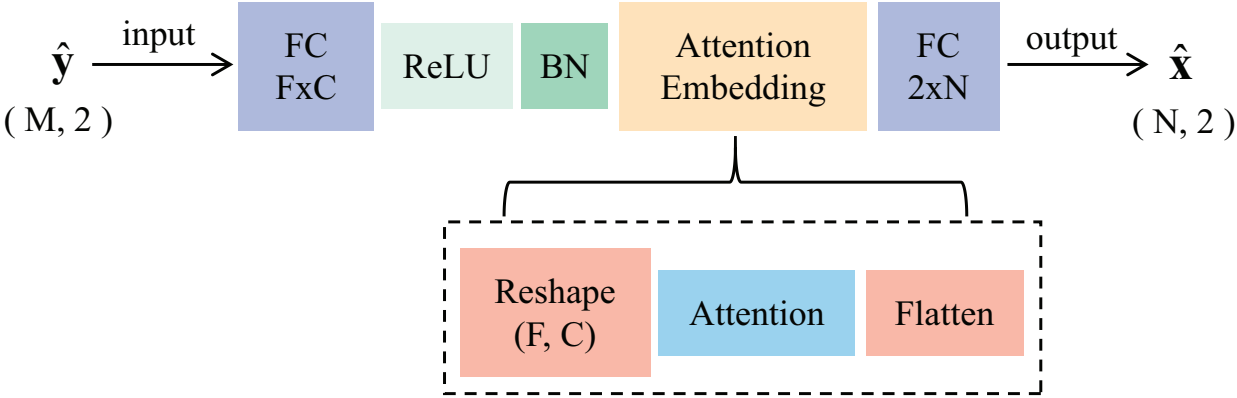}
\caption{Channel estimation network structure with HAD.}
\label{net2}
\end{figure}

As introduced earlier, the attention module is inserted after a feature matrix and the attention map is learned from the squeezed feature matrix obtained by global average pooling. FNN can not directly use attention since all the neurons of the neighboring FC layers are fully connected and features of FC layers appear in the form of vectors instead of matrices. Therefore, as depicted in the dashed box in Fig. \ref{net2}, we reshape the feature vector of a FC layer into a matrix first, like the feature matrix of a Conv1D layer. Then, with the matrix-shaped feature, the original attention mechanism can be normally applied. Finally, the reweighted feature vector can be obtained by flattening the reweighted feature matrix.

The detailed network design is illustrated in Fig. \ref{net2}. The first FC layer consists of $F\times C$ neurons, which is followed by a ReLU activation and a BN layer. The feature vector is then reshaped into a $(F,C)$ feature matrix, where $C$ and $F$ can be regarded as the number of channels and the number of features of each channel, respectively. Based on the feature matrix, the original attention module is inserted to get the reweighted feature matrix, which is then flattened back to the reweighted feature vector. Eventually, an output FC layer with $2N$ neurons is used to obtain the real and imaginary parts of the angular domain channel prediction. We only use one hidden FC layer here since experiments indicate that more hidden FC layers are not helpful to further improve the performance but increases the complexity dramatically.

\subsection{Complexity Analysis}
In this subsection, the complexity of various algorithms are analyzed. Two metrics are used to measure the complexity, namely the required number of floating point operations (FLOPs) and the total number of parameters. For brevity, only multiplication is considered and one complex multiplication is counted as four real multiplications when computing FLOPs, and the weights and biases of BN layers are ignored and one complex parameter is counted as two real parameters when computing parameter number. When analyzing the complexity of neural networks, we ignore the offline training phase and focus on the online testing phase since the network training only needs to be executed once and the BS usually has sufficient computational ability in practice.

Using the notations in Section III, the FLOPs of the Conv1D layer and the $l$-th FC layer are $LFCC^\prime$ and $N_{l-1}N_l$, respectively, where $N_{l}$ denotes the number of neurons in the $l$-th FC layer. Without HAD, the overall FLOPs of CNN can be readily obtained as $(2L_IF+2L_OF+L_HF^2(N_B-1))N$ and the additional FLOPs of attention modules is $N_BF(N+F+1)$. The FLOPs of MMSE estimation is $4(2N^3+N^2)$. Besides, for both algorithms, the LS estimation has to be obtained first, which also requires $4NL_p^2$ FLOPs. In the scenario with HAD, the FLOPs of the attention-aided FNN can be obtained as $FC(2M+2N+1)+C(C+1)$. In this paper, structured variational Bayesian inference (S-VBI) is selected as the CS-based baseline algorithm, whose FLOPs is $I_E(\frac{2}{3}M^3+(2M+2)N^2)$ with $I_E$ denoting the number of iterations\cite{VBI}. Again, for both algorithms, obtaining the received signal corresponding to a single user requires $4ML_p^2$ FLOPs. Notice that in both scenarios, the FLOPs of DL-based algorithms only scale linearly with $N$ and $M$, which is an attractive practical advantage, especially in large scale systems. By contrast, the FLOPs of conventional algorithms are much higher and grow cubically with $N$ and $M$.

As for the total number of parameters, the Conv1D layer and the $l$-th FC layer contains $LCC^\prime$ and $N_{l-1}N_l$ parameters, respectively. Without HAD, CNN contains totally $2(L_I+L_0)F+L_H(N_B-1)F^2$ parameters and the additional number of parameters of attention modules is $N_BF^2$. The CCM used in MMSE requires $2N^2$ parameters. In the scenario with HAD, attention-aided FNN contains totally $FC(2M+2N)+C^2$ parameters, while S-VBI does not need any parameters.

\section{Simulation Results}
In this section, extensive simulation results are presented to evaluate the performance of the proposed DL-based channel estimation framework in scenarios with and without HAD. MSE is adopted as the performance metric. Notice that, converting the channel to angular domain does not change the MSE since $\bm{F}$ is a unitary matrix. Some of the parameters used in simulation are summarized in Table \ref{parameters}, unless otherwise specified. As for network hyper-parameters in the scenario without HAD, $L_I$, $L_H$, and $L_O$ are set to 7, 5, and 1, respectively, and $S$ is set to 1.
\begin{table}[!htbp]
\centering
\begin{tabular}{|c|c|}
\hline
Parameter & Value \\
\hline
$N$ & 128\\
\hline
$M$ & 32\\
\hline
$N_p$ & 20\\
\hline
$\theta_i$ & $\mathcal{U}[0,2\pi]$\\
\hline
$\bigtriangleup_{\theta}$ & $5^\circ$ \\
\hline
$\alpha_i$ & $\mathcal{CN}(0,1)$\\
\hline
SNR & 20 dB \\
\hline
\end{tabular}
\caption{Simulation parameters.}
\label{parameters}
\end{table}

We compare the proposed algorithm with the following baseline algorithms. The structures of all DL-based baselines are carefully determined by cross validation out of fairness.

\subsubsection{Without HAD}
The following algorithms are selected as baselines:
\begin{itemize}
\item{\bfseries MMSE Single:} Refine the LS estimation by the CCM, $\bm{R}_{hh}\triangleq \mathbb{E}({\bm{h}\bm{h}^H})\in \mathbb{C}^{N \times N}$ as \cite{LS_MMSE}
\begin{equation}
\hat{\bm{h}}_{\text{MMSE}}=\bm{R}_{hh}(\bm{R}_{hh}+\bm{I}/\text{SNR})^{-1}\hat{\bm{h}}_{\text{LS}}.
\end{equation}
\item{\bfseries MMSE $\boldsymbol{3^\circ$}:} Split the entire angular space into many $3^\circ$-angular regions and estimate a dedicated CCM for each region with only channel samples whose average AoAs are in the region. During the testing process of a channel sample, the angular region it belongs to will be estimated first\footnote{The angular region estimations of samples are assumed to be accurate for simplicity.} and the corresponding CCM will be selected for channel refinement. Compared with using a single CCM for all channel samples, using multiple CCMs matching different angular regions can effectively exploit the narrow angular spread characteristic of channels and improve performance significantly. Actually, it can be regarded as the manual implementation of the ``divide-and-conquer" policy, i.e., the channel samples are ``divided" by their angular regions and ``conquered" by different corresponding CCMs. 
\item{\bfseries FNN:} The FNN structure consists of three FC layers with 512, 1024, and 256 neurons, respectively, with one BN layer inserted between every two FC layers. The activation function of the first two FC layers is ReLU while the last FC layer does not use activation.
\item{\bfseries CNN without attention:} The same CNN structure but with all the attention modules removed.
\end{itemize}

\subsubsection{With HAD}The following algorithms are selected as baselines:
\begin{itemize}
\item{\bfseries Separate LS:} A total of $N/M$ estimates are executed. In each estimate, only $M$ antennas are switched on by adjusting $\bm{W}$, and their channels are obtained by LS estimation\cite{HAD_angular}.
\item{\bfseries S-VBI:} One of the state-of-the-art CS-based algorithms designed for narrow angular spread channel estimation in HAD massive MIMO systems, where the spatial-clustered channel sparsity is embedded to improve the estimation performance\cite{VBI}. The source code is provided by the authors of \cite{VBI}.
\item{\bfseries FNN without attention:} Adopt the same structure of FNN as in the former scenario while the number of neurons reduces to 256, 512, and 256, respectively, with smaller input dimension.
\item{\bfseries CNN:} The structure of CNN is also similar to the former scenario, except that the output layer is changed from Conv1D to FC for dimension conversion.
\item{\bfseries CNN without attention:} The same CNN structure but with all the attention modules removed.
\end{itemize}

\subsection{Impacts of Network Parameters}
\begin{figure}[htbp] 
	\centering  
	\vspace{-0.35cm} 
	\subfigtopskip=2pt 
	\subfigbottomskip=1pt 
    \subfigcapskip=-3pt 
	\subfigure[Without HAD]{
		\label{impact_of_scale_CE}
		\includegraphics[width=0.48\textwidth]{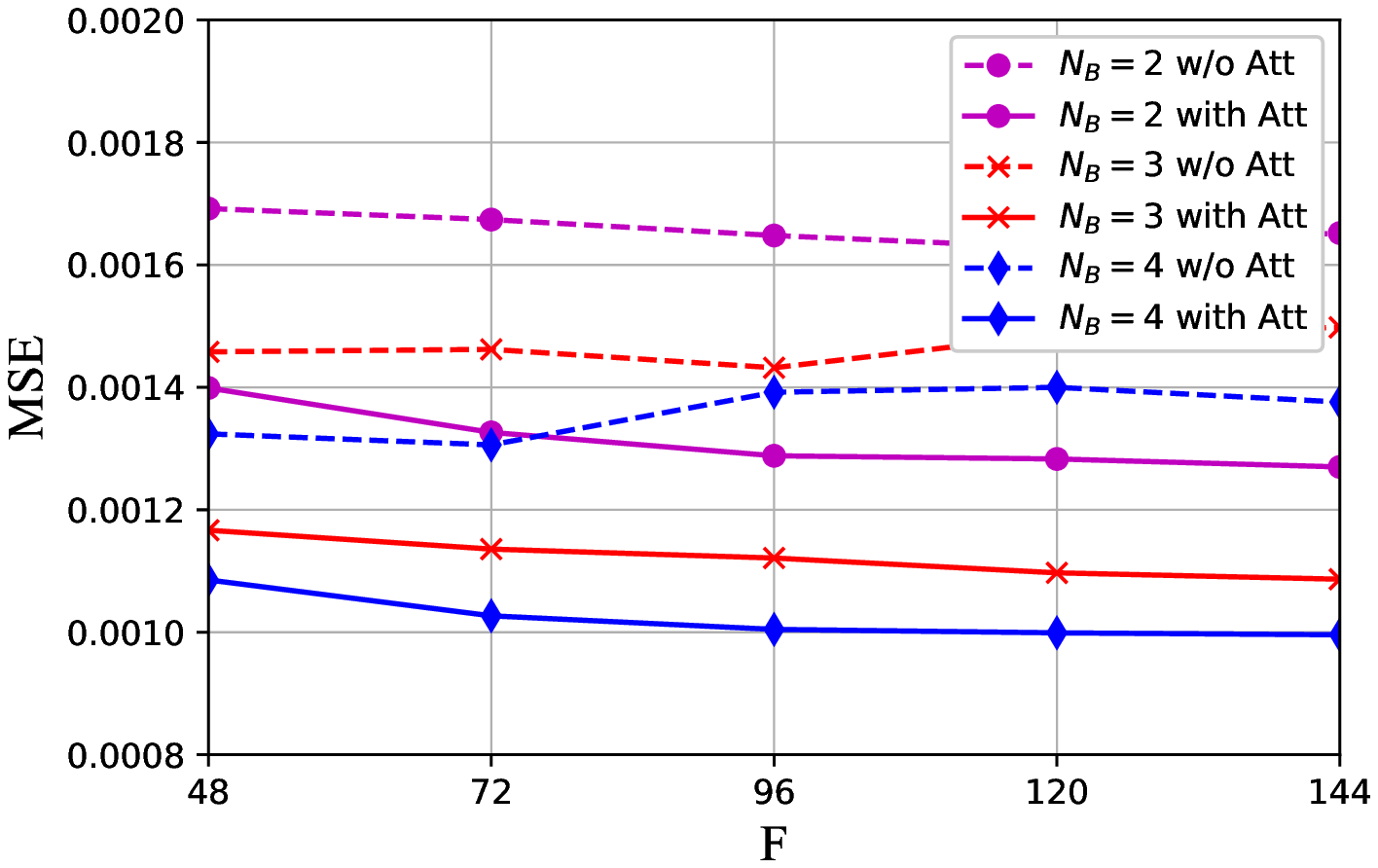}}
	\subfigure[With HAD]{
		\label{impact_of_scale_CS}
		\includegraphics[width=0.48\textwidth]{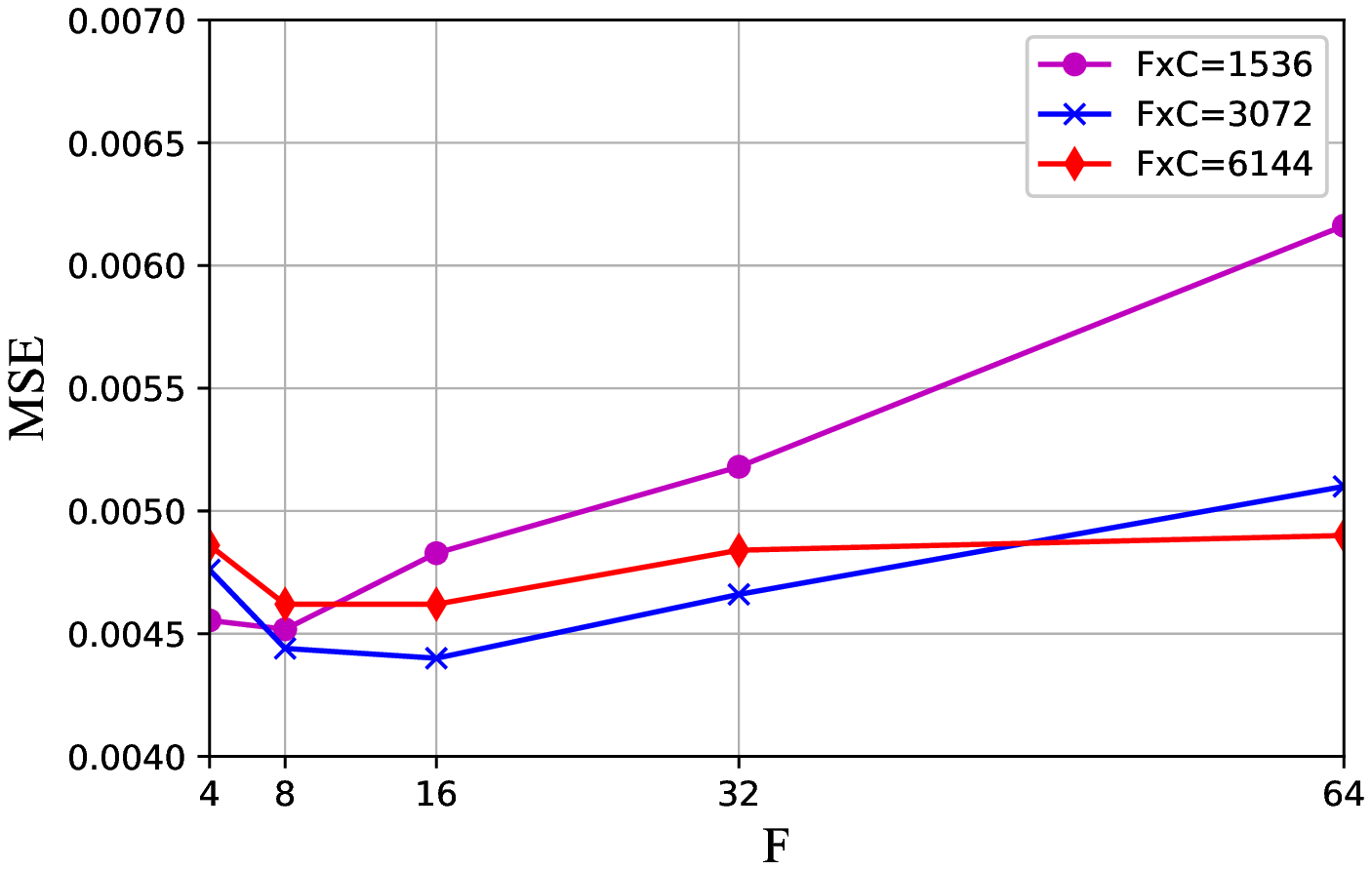}}
	\caption{Impact of network parameters in the two considered scenarios.}
	\label{impact_of_scale}
\end{figure}

To determine the best network structures for two scenarios, we investigate the impacts of key network parameters on network performance. Without HAD, the structure of CNN is mainly determined by the number of convolutional blocks, $N_B$, and the number of filters of each Conv1D layer, $F$. As illustrated in Fig. \ref{impact_of_scale_CE}, attention can improve the performance of CNNs with various numbers of convolutional blocks and filters and the performance of a two-layer attention-aided CNN is even better than a four-layer CNN without attention, which indicates the superiority of the attention mechanism. In general, the performance of networks is better with stronger representation capability brought by more convolutional blocks. However, with enough filters, the performance improvement of attention-aided CNN is marginal if the number of filters keeps growing and it can even be harmful to CNN without attention sometimes. Besides, deeper and wider CNNs also have heavier computing and storage burdens. To strike a balance between performance and complexity, we choose to use four convolutional blocks and 96 filters for each Conv1D layer.

With HAD, the structure of the attention-aided FNN is mainly determined by the number of neurons of the hidden FC layer $F\times C$ and the way of reshaping in the attention embedding module. As in Fig. \ref{impact_of_scale_CS}, the network performs best when $F\times C=3072$ and the performance will deteriorate with either too few or many neurons. Besides, as can be indicated from the bowl shape of curves, a medium number of features in each channel performs best when $F\times C$ is fixed. The reason is that the number of channels is too small and there is not enough degrees of freedom for dynamic adjustment of attention maps when $F$ is too large, while each channel does not contain enough features to effectively capture the global information\cite{attention} when $F$ is too small. So, we choose to reshape the feature vector into 192 channels with 16 features in each channel.

\subsection{Impacts of System Parameters}
In this subsection, the impacts of various system parameters are investigated to validate the superiority and universality of the proposed approach.

\subsubsection{Impact of SNR}
\begin{figure}[htbp] 
	\centering  
	\vspace{-0.35cm} 
	\subfigtopskip=2pt 
	\subfigbottomskip=1pt 
    \subfigcapskip=-3pt 
	\subfigure[Without HAD]{
		\label{impact_of_snr_CE}
		\includegraphics[width=0.48\textwidth]{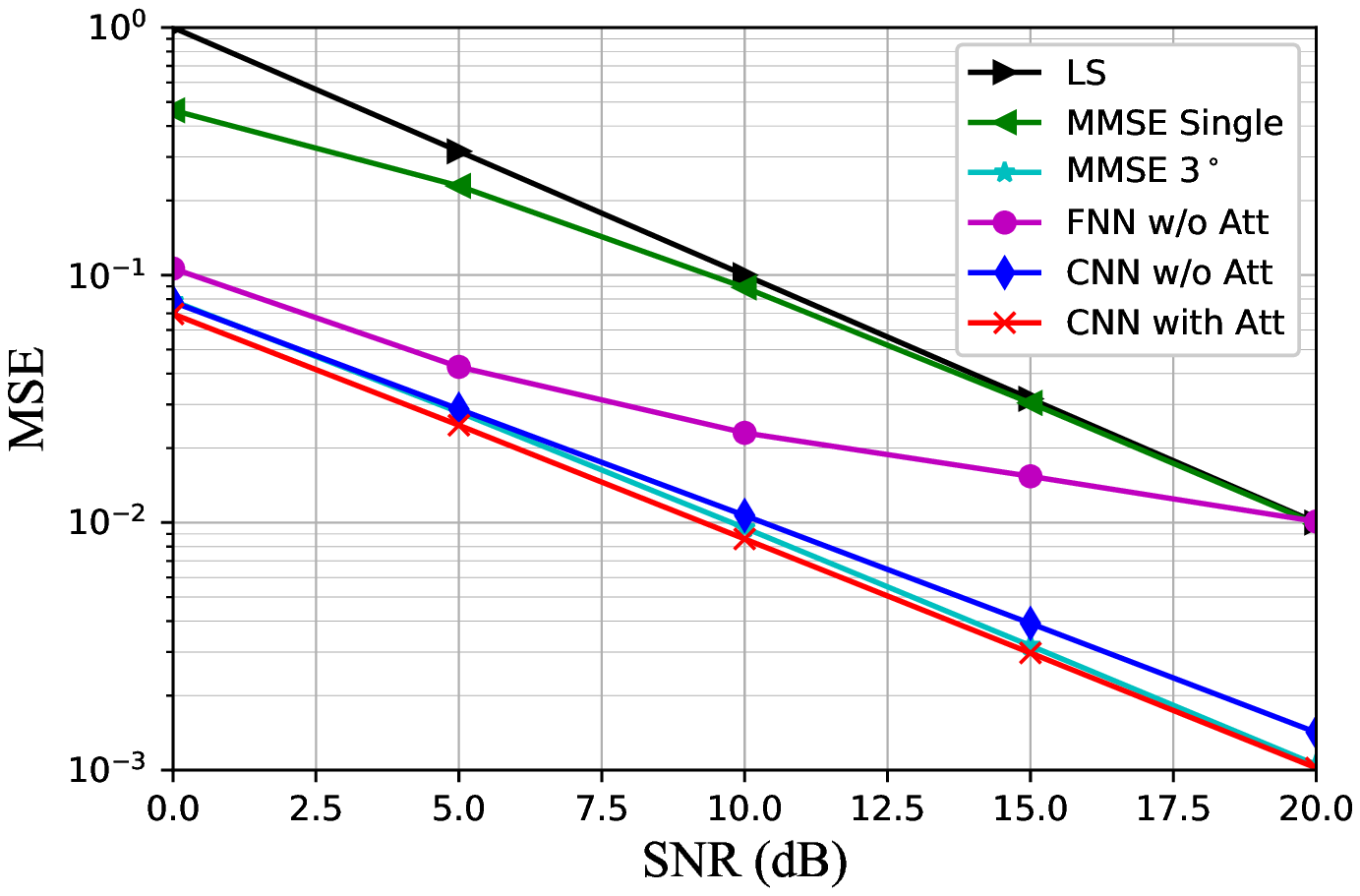}}
	\subfigure[With HAD]{
		\label{impact_of_snr_CS}
		\includegraphics[width=0.48\textwidth]{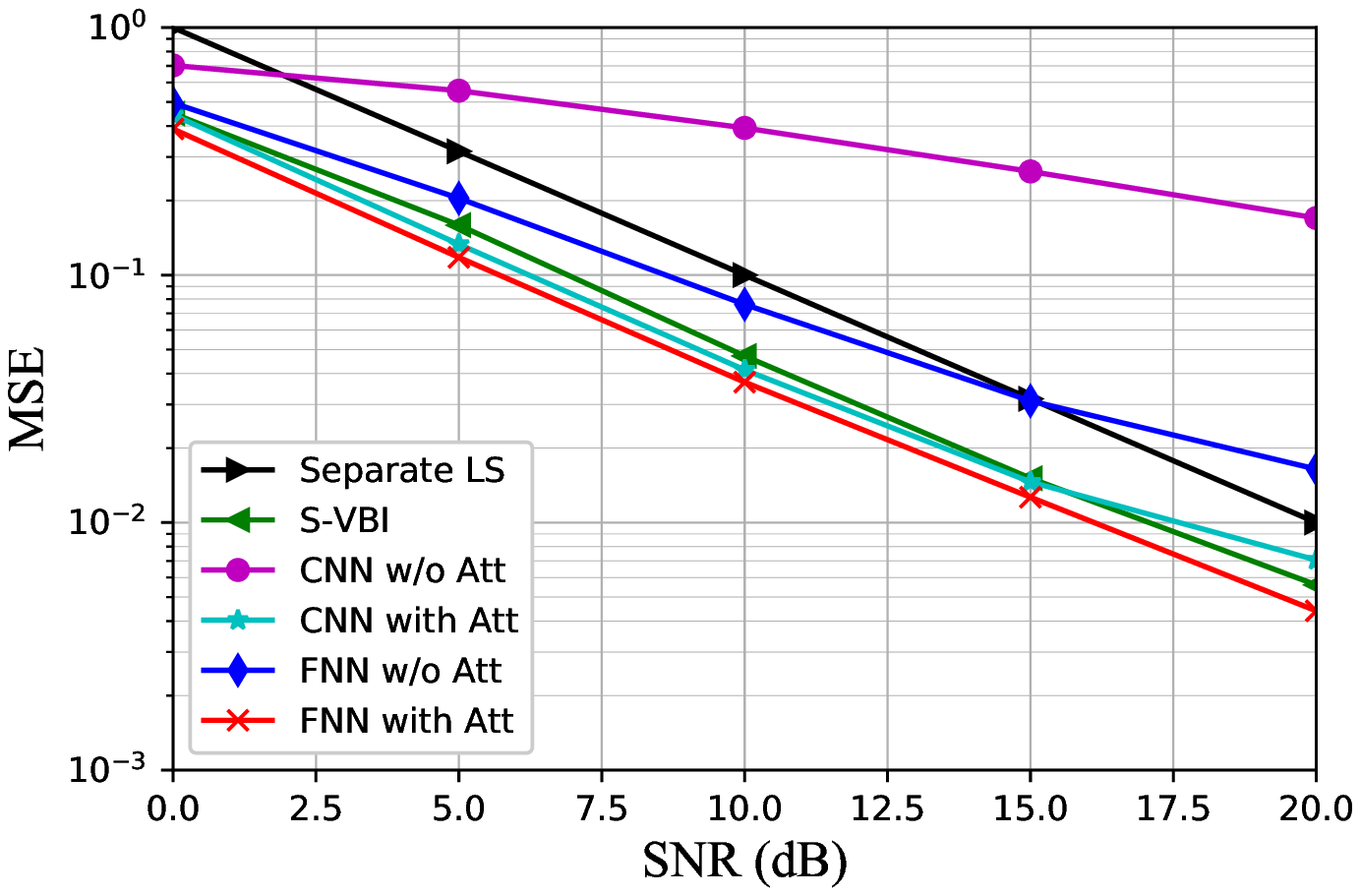}}
	\caption{Impact of SNR in the two considered scenarios.}
	\label{impact_of_snr}
\end{figure}

As illustrated in Fig. \ref{impact_of_snr_CE}, without HAD, all DL-based methods can refine and improve the channel quality of LS coarse estimation. The performance improvement of FNN decreases as the SNR increases while CNN outperforms LS significantly in various SNR regimes thanks to the exploitation of local correlation of input data. Then, with the aid of attention, the MSE of CNN further decreases moderately. Besides, the performance gain of the attention mechanism increases with SNR. When SNR is 0 dB, the MSE of CNN with attention is $89.55\%$ of that of CNN without attention while this ratio decreases to $71.83\%$ when SNR is 20 dB. The reason is that the narrow angular spread characteristic of the angular domain channel is more exposed and easier to be exploited with less noise, thereby amplifying the benefits of attention. As for MMSE, the performance improvement of MMSE Single is marginal while MMSE $3^\circ$ performs much better due to the exploitation of the narrow angular spread characteristic of channel. Nevertheless, the proposed attention-aided CNN still slightly outperforms MMSE $3^\circ$, demonstrating its superiority.

From Fig. \ref{impact_of_snr_CS}, the performance of FNN is much better than CNN and outperforms separate LS except in high SNR regimes when HAD is considered and attention is not used, but it is still obviously inferior to S-VBI. However, with the aid of attention, the performance of both CNN and FNN improves significantly. As can be observed, attention-aided CNN outperforms S-VBI except when SNR is higher than 15 dB while the attention-aided FNN is even better and outperforms S-VBI consistently in all SNR regimes. Besides, compared with Fig. \ref{impact_of_snr_CE}, the performance gain of the attention mechanism is much more significant since the attention mechanism can not only help denoise but also plays an important role in reversing the effect of $\bm{W}$ in the HAD scenario. Specifically, when restoring the high-dimensional channel from the low-dimensional received signal, the performance deterioration can be effectively reduced if the approximate AoA range of channel paths is known. Thanks to the attention mechanism, such processing can be automatically realized by the dynamic adjustment of attention maps.

\subsubsection{Impact of Angular Spread}
\begin{figure}[htbp] 
	\centering  
	\vspace{-0.35cm} 
	\subfigtopskip=2pt 
	\subfigbottomskip=1pt 
    \subfigcapskip=-3pt 
	\subfigure[Without HAD]{
		\label{impact_of_AS_CE}
		\includegraphics[width=0.48\textwidth]{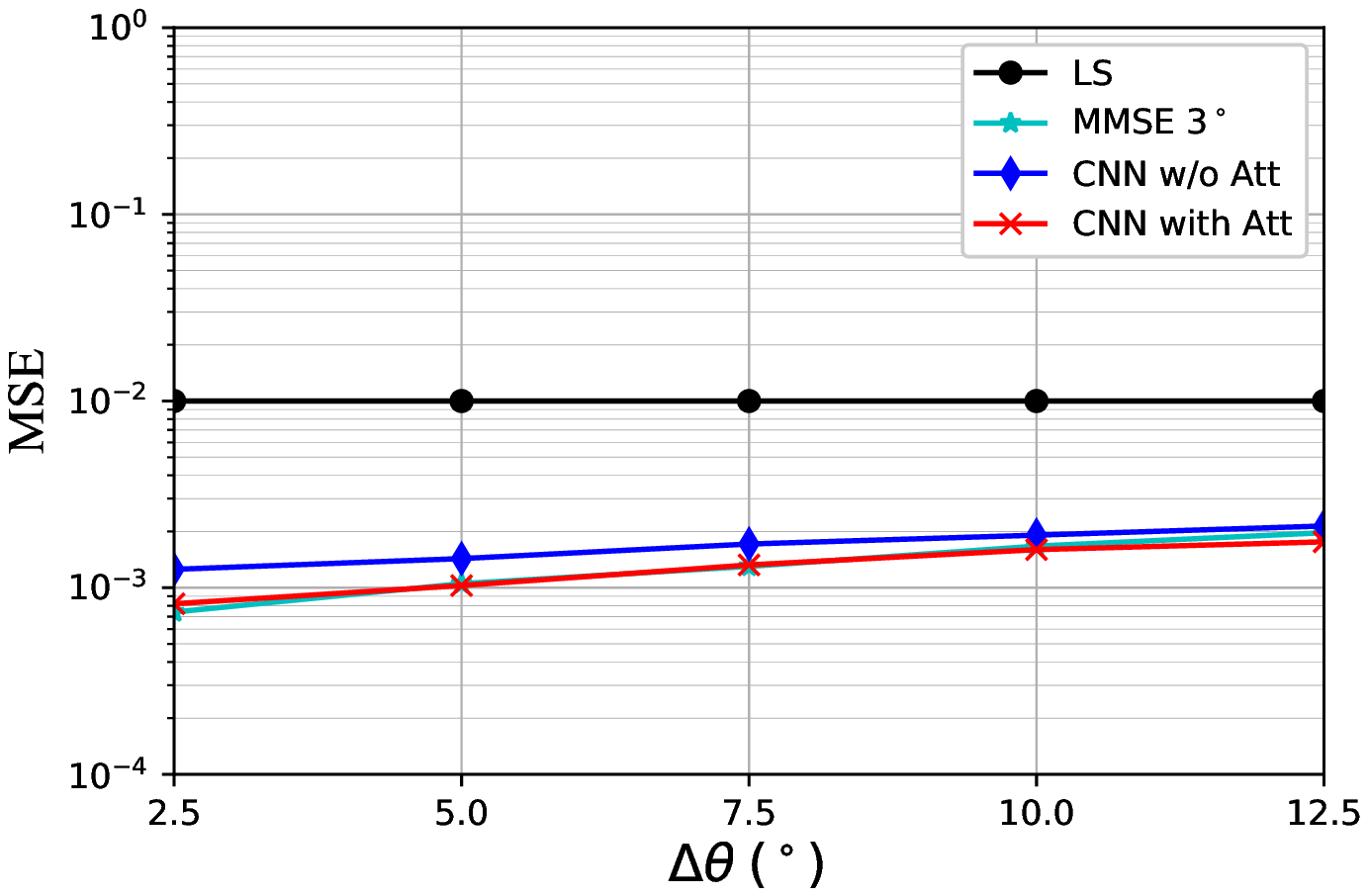}}
	\subfigure[With HAD]{
		\label{impact_of_AS_CS}
		\includegraphics[width=0.48\textwidth]{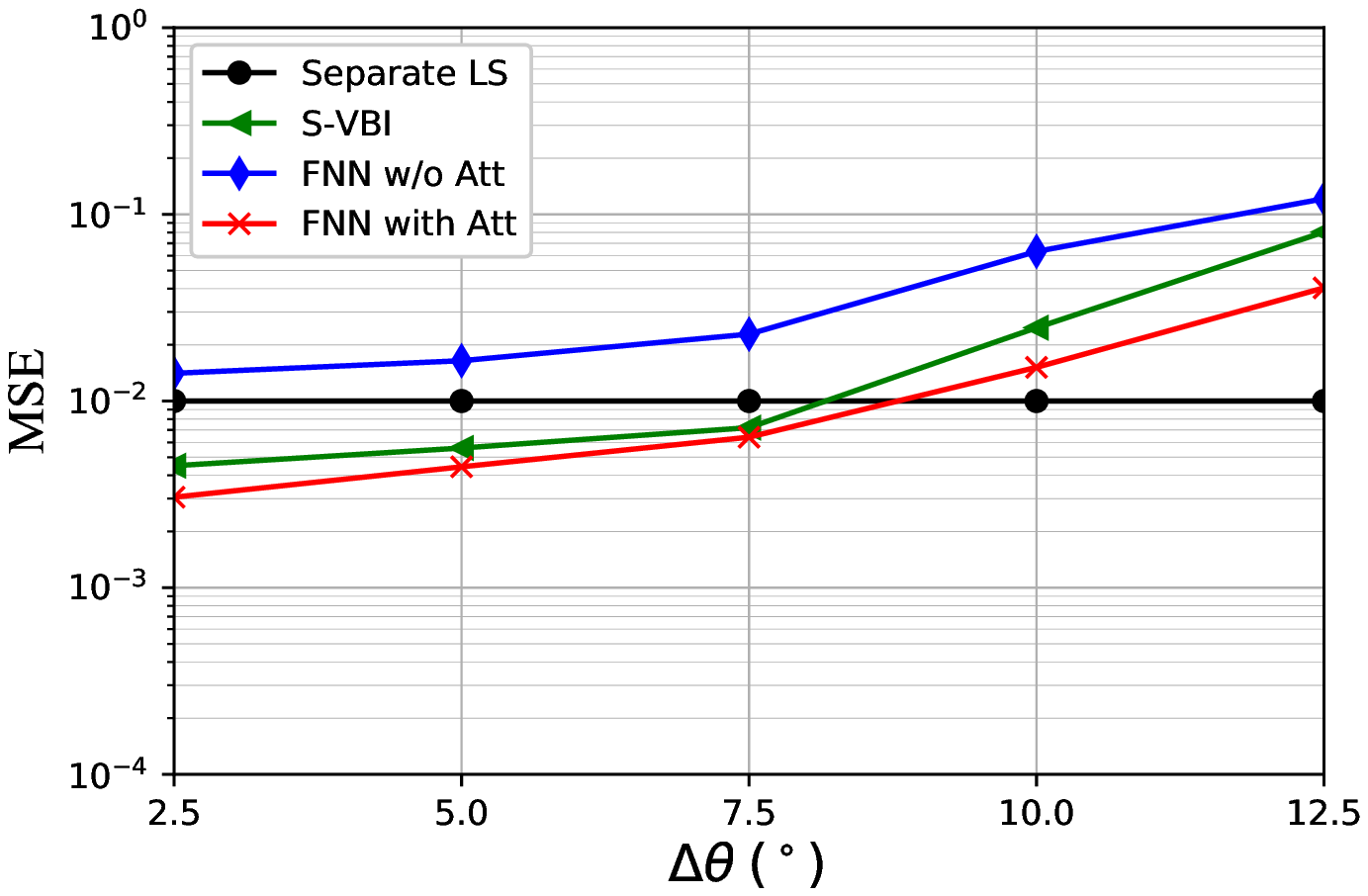}}
	\caption{Impact of angular spread in the two considered scenarios.}
	\label{impact_of_AS}
\end{figure}

As is illustrated in Fig. \ref{impact_of_AS}, attention-aided CNN has close performance to MMSE $3^\circ$ and consistently outperforms LS significantly with various angular spreads. As angular spread increases, the performance of all algorithms decreases in both scenarios since the channel estimation problem becomes more complex with less sparse angular domain channel. Besides, the performance gain of attention also decreases because the channel distribution is less separable, which makes the attention mechanism more difficult to realize the ``divide-and-conquer" policy. In the scenario with HAD, the performance of attention-aided FNN is better than separate LS unless the angular spread is too large while only $M/N$ resource overhead is required.

\subsubsection{Impacts of Antenna Number and RF Chain Ratio}
\begin{figure}[htbp] 
	\centering  
	\vspace{-0.35cm} 
	\subfigtopskip=2pt 
	\subfigbottomskip=1pt 
    \subfigcapskip=-3pt 
	\subfigure[Without HAD]{
		\label{impact_of_Nt_CE}
		\includegraphics[width=0.48\textwidth]{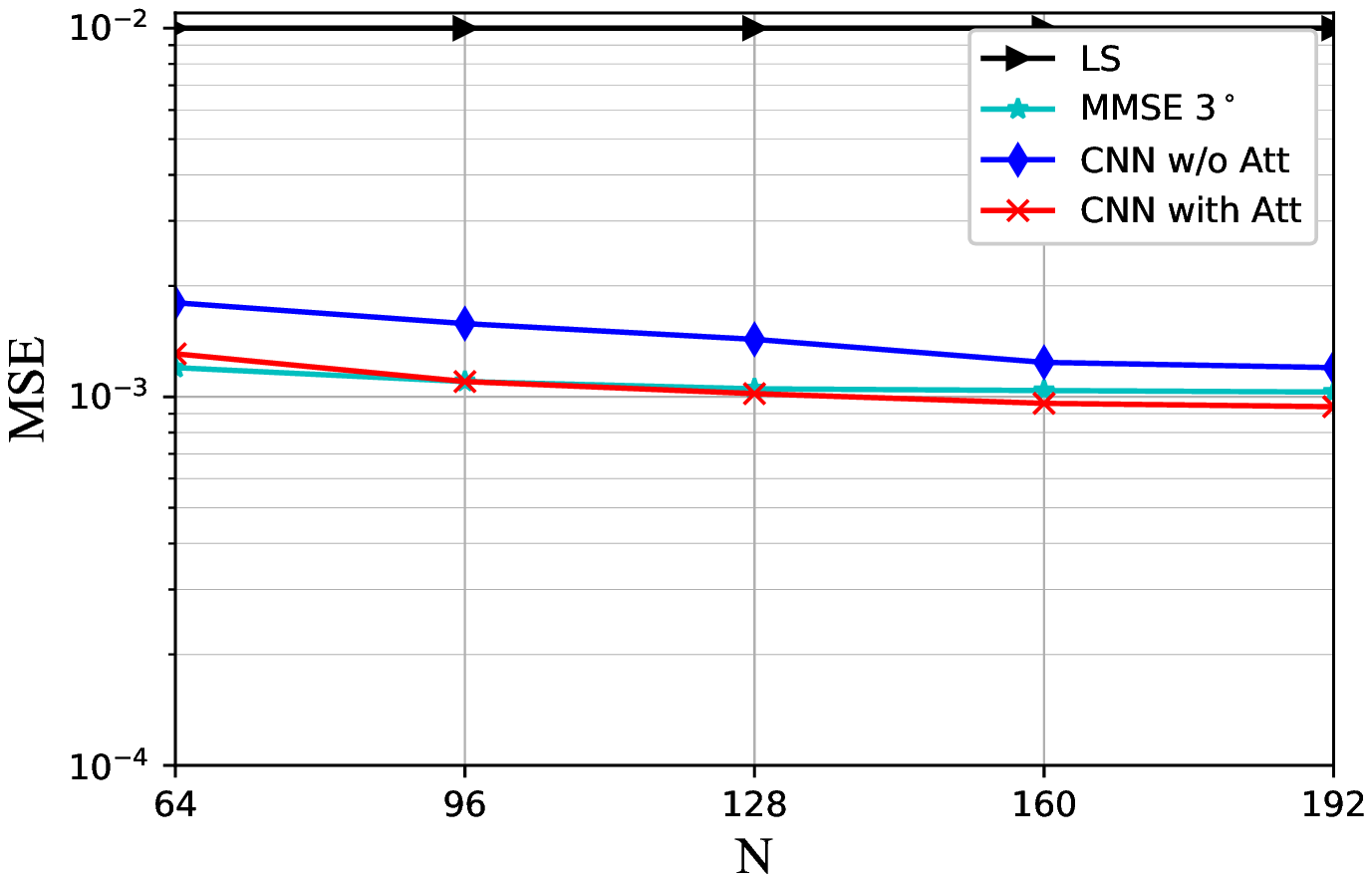}}
	\subfigure[With HAD]{
		\label{impact_of_Nt_and_RF_CS}
		\includegraphics[width=0.48\textwidth]{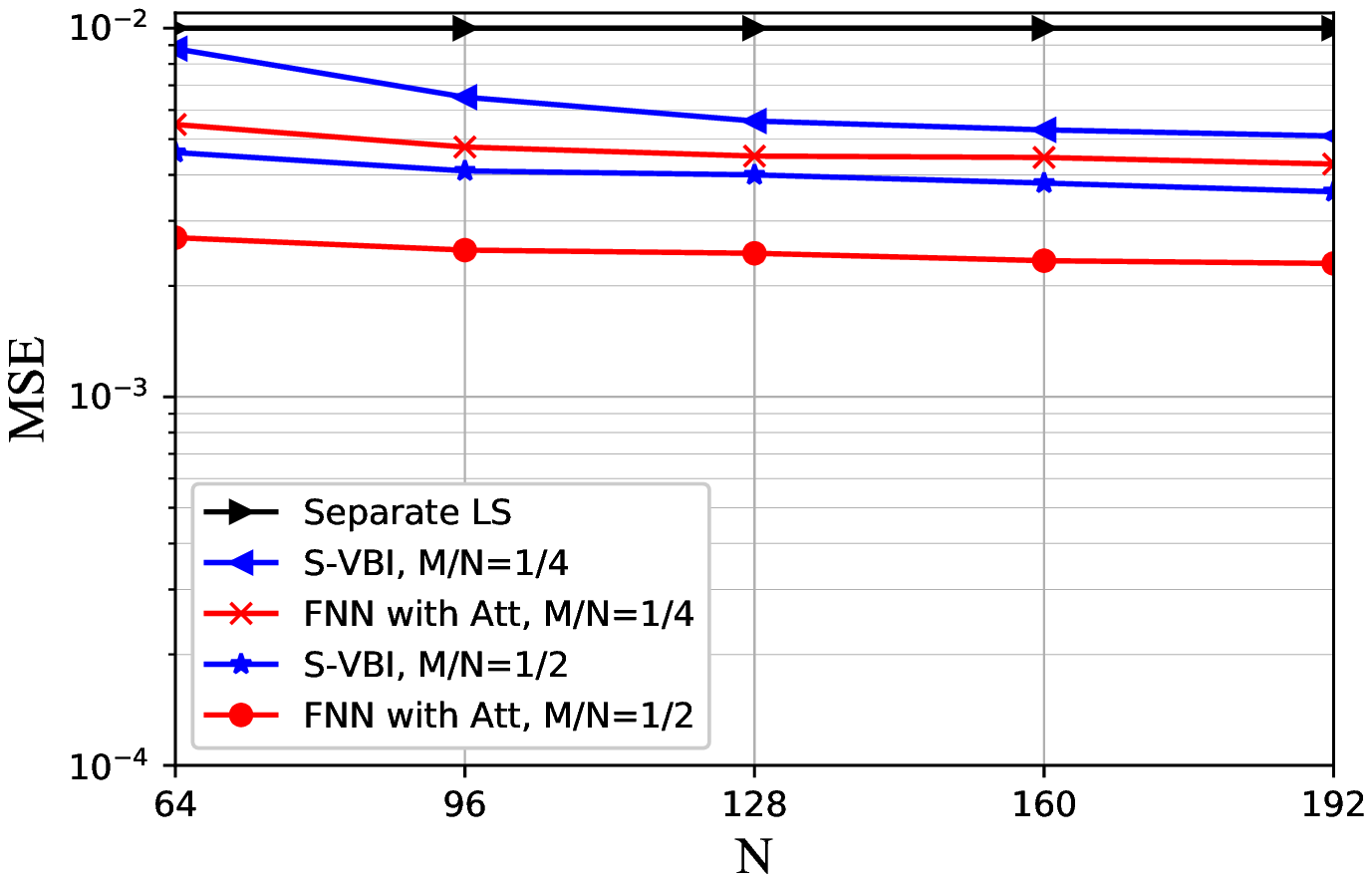}}
	\caption{Impact of antenna number and RF chain ratio in the two considered scenarios.}
	\label{impact_of_Nt_and_RF}
\end{figure}

As can be observed from Fig. \ref{impact_of_Nt_and_RF}, in both scenarios, the performance of all algorithms improves as $N$ increases. Since the power leakage of angular domain channel is inversely proportional to the antenna number\cite{Unified}, the increased channel sparsity caused by more antennas can simplify channel estimation. Without HAD, attention-aided CNN has close performance to MMSE $3^\circ$ and the performance gain of attention can be amplified by sparser channel. With HAD, the performance of all algorithms improve as the RF chain ratio $M/N$ increases since more information is kept during the sensing phase. Besides, attention-aided FNN outperforms S-VBI consistently with various $M$ and $N$ and the performance gap increases with less antennas with fixed RF chain ratio, indicating that the DL-based approach is less dependent on channel sparsity. From the perspective of resource saving, attention-aided FNN is also superior to S-VBI. In particular, the MSE of attention-aided FNN with only $1/4$ RF chains is comparable to that of S-VBI with $1/2$ RF chains. As a result, the hardware and energy cost can be halved. Furthermore, given strict target MSE performance and a limited number of RF chains, S-VBI may need to estimate multiple times while attention-aided FNN completes the estimation at once, saving more resources for data transmission. Such an advantage can be very appealing in scenarios like high-mobility communication, where the channel is fast time-varying with short channel coherence time.

\subsection{{Generalization Ability}}
The generalization ability to different parameters heavily influences the practicality of neural networks. In the considered problem, there are two categories of parameters, namely system parameters and channel parameters. System parameters include the number of antennas, RF chains, and users, which determine the input and output dimensions of the network. Channel parameters include SNR, number of paths, angular spread, and gain distribution of channel paths, which influence the input and output distributions of the network. For system parameters, the numbers of antennas and RF chains are usually fixed in practice, and different user numbers can also be handled by the same network since a multi-user channel estimation problem is decomposed into multiple single-user problems by exploiting the orthogonality of pilot sequences. Therefore, we focus on the generalization performance of channel parameters.

\begin{figure}[htb!] 
	\centering  
	\vspace{-0.35cm} 
	\subfigtopskip=2pt 
	\subfigbottomskip=1pt 
    \subfigcapskip=-3pt 
	\subfigure[SNR generalization without HAD.]{
		\label{snr_gen_ce}
		\includegraphics[width=0.48\textwidth]{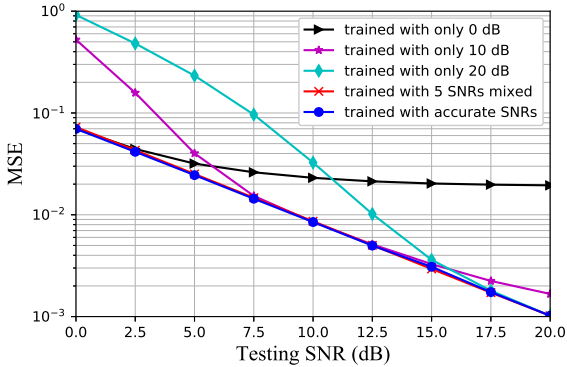}}
	\subfigure[SNR generalization with HAD.]{
		\label{snr_gen_cs}
		\includegraphics[width=0.48\textwidth]{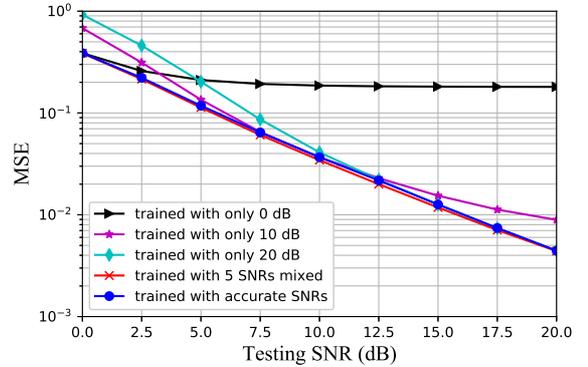}}
	\caption{Generalization to SNRs with different training methods in the two considered scenarios.}
	\label{snr_gen}
\end{figure}

The generalization to different SNRs is illustrated in Fig. \ref{snr_gen}. The legend ``trained with accurate SNRs" denotes that for each SNR, a dedicated model trained with accurate SNR data is used for testing. In both scenarios, the proposed networks can only handle tiny SNR mismatch between the training and testing phases when the model is trained with a single SNR point and the performance degradation can be very severe when the SNR mismatch is large. To alleviate this issue, one common method is training with data under a variety of SNRs, then the characteristics of different SNRs can be captured by a single network. In simulation, we select five SNR points, namely 0, 5, 10, 15, and 20 dB for training. Besides, the number of training samples from each SNR point is kept same as when trained separately out of fairness. Based on our simulation results, directly using MSE as loss can lead to poor performance when different SNR points are trained together since the loss of high SNR data will be overwhelmed by the loss of low SNR data. To ensure that all SNR regimes get sufficient training, we use a heuristic loss function computed as
\begin{equation}
\text{Weighted MSE Loss}=\frac{1}{n}\sum_{i=1}^{n}(\text{SNR}_i\cdot \left\|\hat{\bm{x}}_i-\bm{x}_i\right\|^2),
\label{weighted mse}
\end{equation}
where the MSE is weighted by the SNR of data sample. As can be indicated by the two close curves marked with circle and cross, networks trained with mixed SNRs achieve similar performance as trained with accurate SNRs and significantly outperform networks trained with a single SNR point. 

As for the generalization to other parameters, detailed results are omitted here due to space limitation while the trends and patterns are also similar. In conclusion, through mixed parameters training and proper design of the loss function, a single network with strong robustness can be obtained to handle all situations during testing, which is very appealing in practical applications.

\subsection{The Role of Attention}
Although it is hard to rigorously analyze the representations learned by DNNs, we still try to attain at least a primitive understanding of the role of attention. Intuitively, the performance gain of attention can be considered to come from the ``divide-and-conquer" policy realized by the dynamic adjustment of attention maps. In this way, sample-specific processing can be performed on different data samples to improve the performance. Without attention, the processing performed by the network is fixed for all data samples, which is less advanced. Next, we would like to analyze the distributions of learned attention maps to roughly corroborate this.

\begin{figure}[htb!] 
	\centering  
	\vspace{-0.35cm} 
	\subfigtopskip=2pt 
	\subfigbottomskip=1pt 
    \subfigcapskip=-3pt 
	\subfigure[The first attention map without HAD]{
		\label{att_map_1}
		\includegraphics[width=0.48\textwidth]{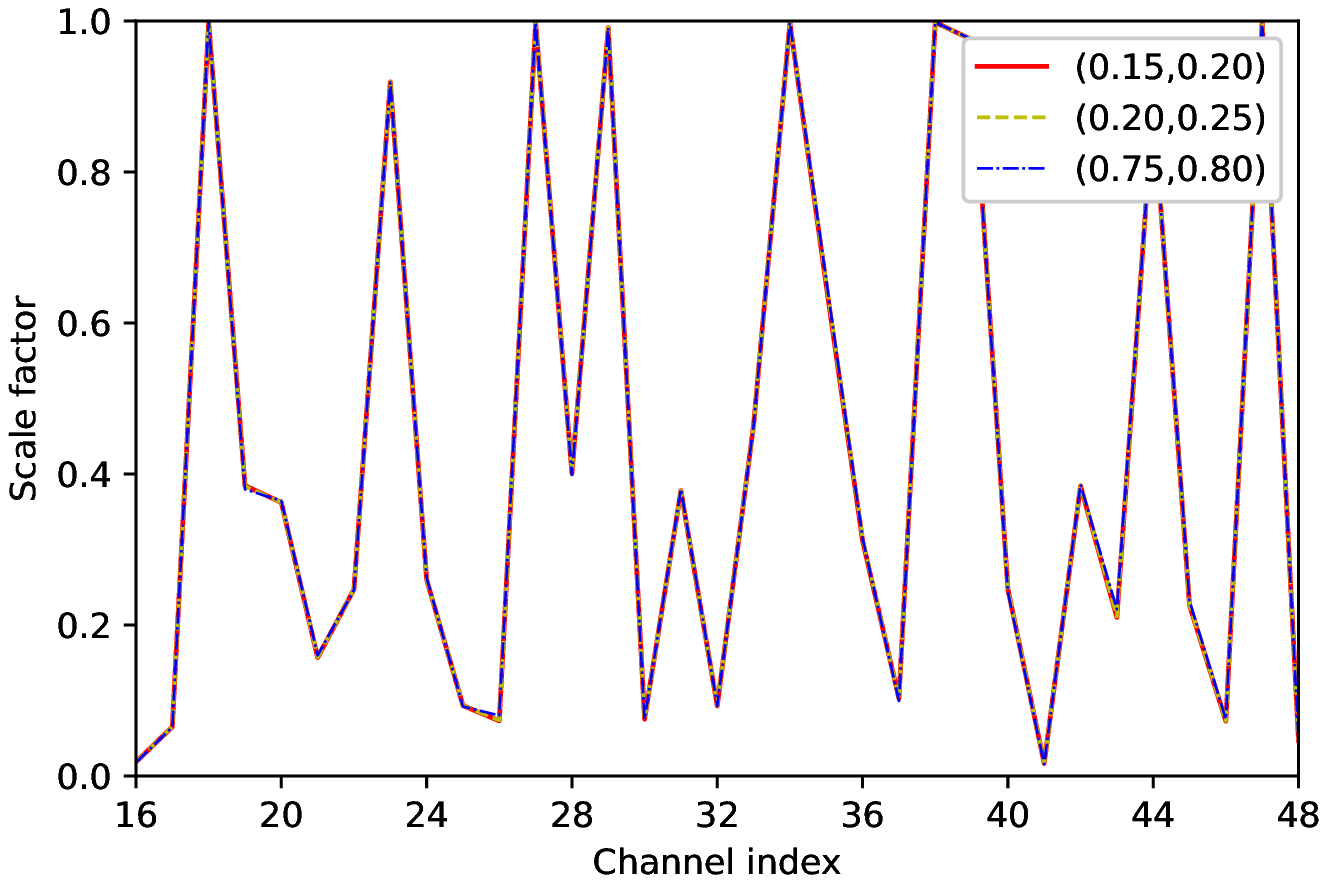}}
	\subfigure[The third attention map without HAD]{
		\label{att_map_3}
		\includegraphics[width=0.48\textwidth]{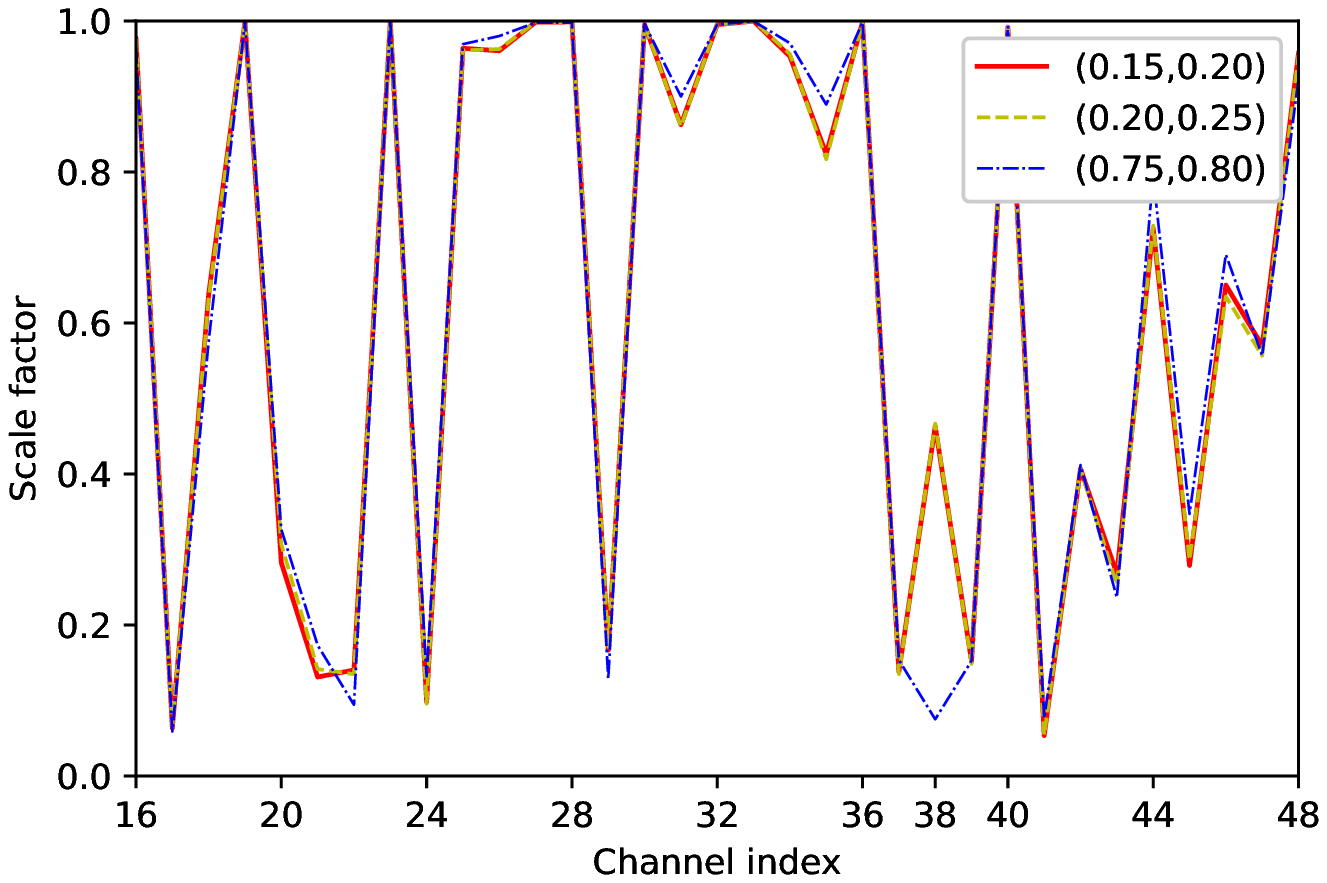}}
	\subfigure[The fourth attention map without HAD]{
		\label{att_map_4}
		\includegraphics[width=0.48\textwidth]{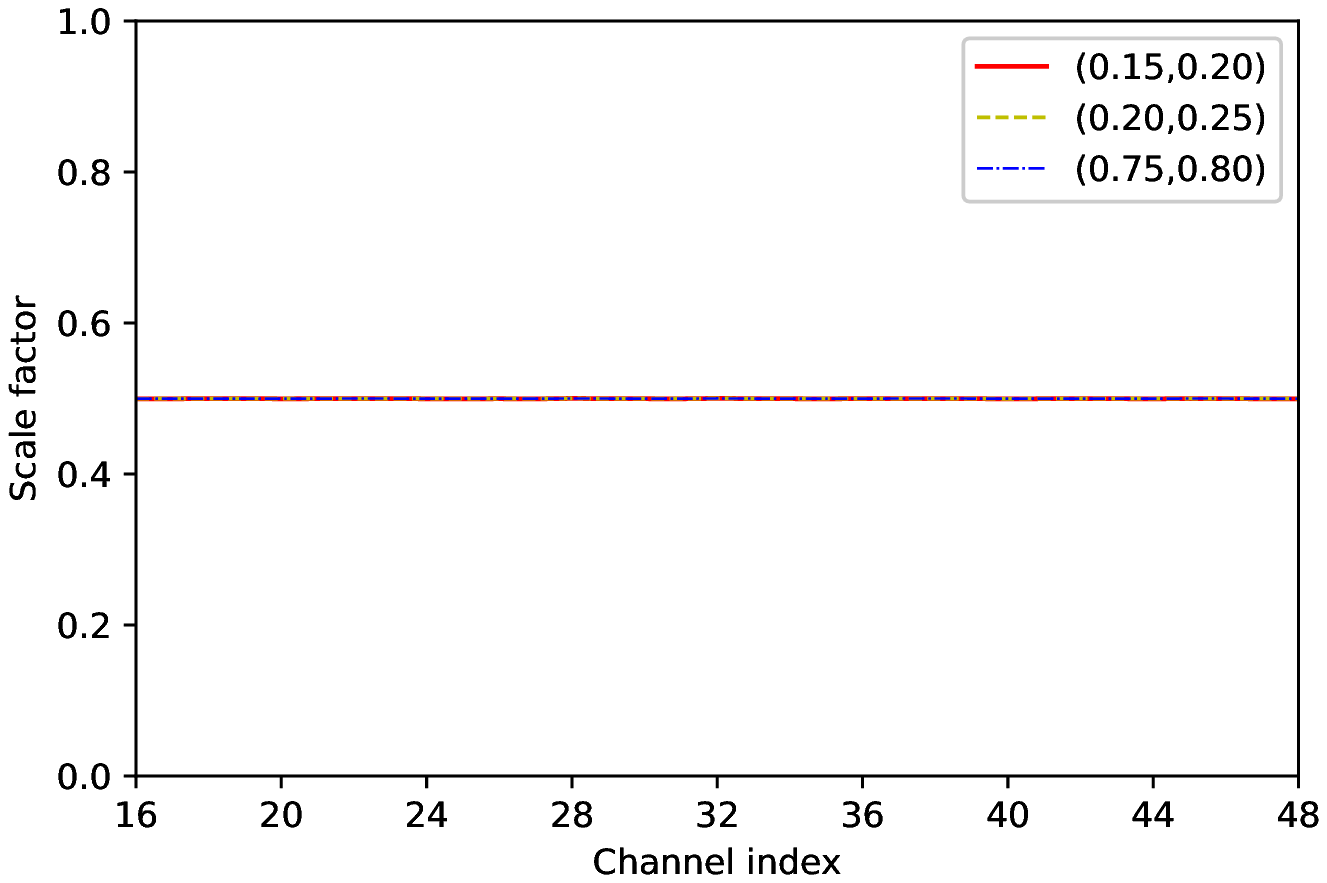}}
	\subfigure[The attention map with HAD]{
		\label{att_map_cs}
		\includegraphics[width=0.48\textwidth]{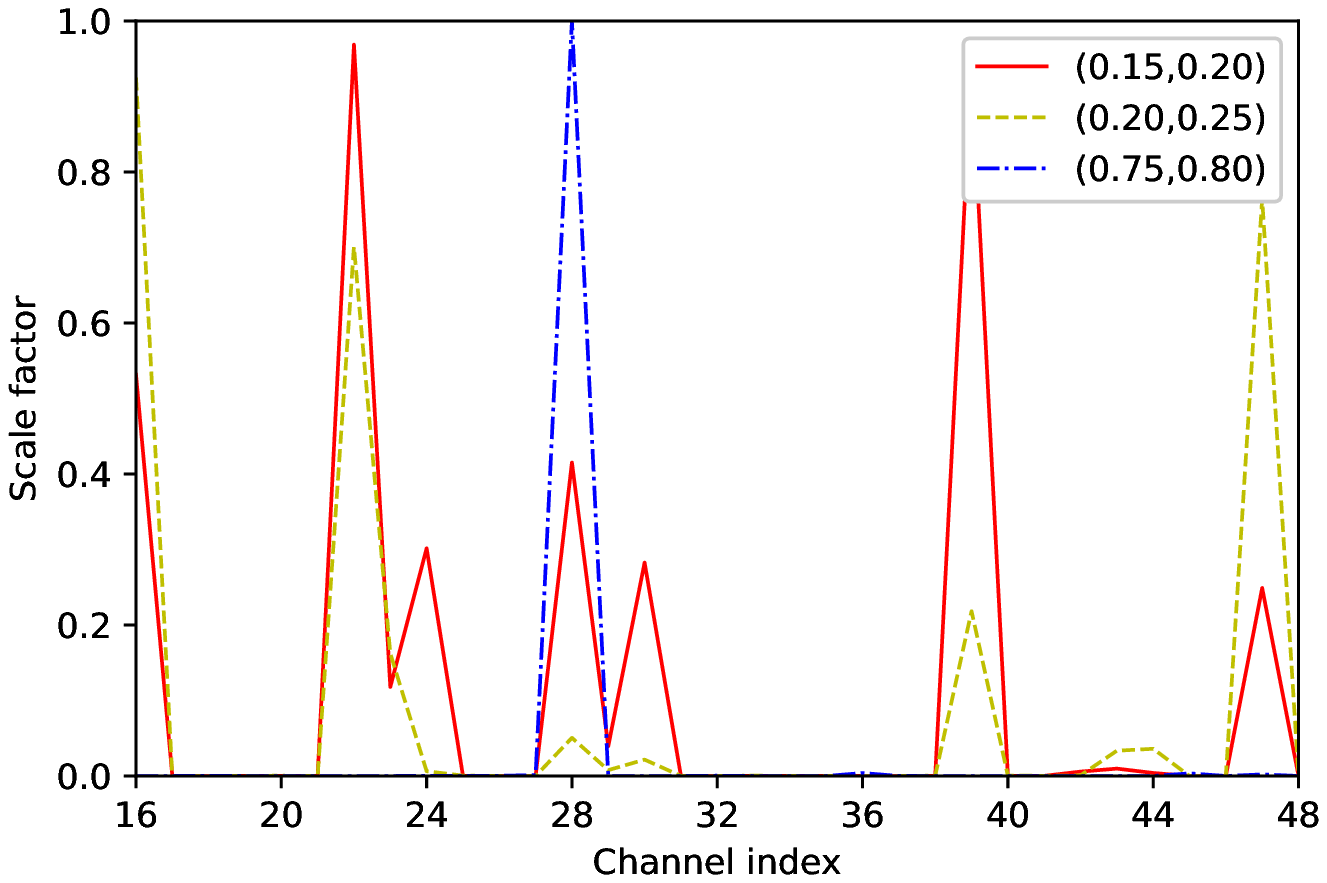}}
	\caption{Average attention maps of data samples in three ranges. The legend $(a,b)$ denotes the range where the minimum and maximum sine values of average AoAs are $a$ and $b$, respectively.}
	\label{CE_att_maps}
\end{figure}

Due to the narrow angular spread characteristic, the channel distribution is highly related to the average AoA parameter, or, more precisely, its sine value. So, we select three sine value ranges for comparison, where the first two ranges are close to each other and the third range is far away from the first two ranges. The average attention maps of validation data samples whose average AoAs are inside the three ranges are plotted in Fig. \ref{CE_att_maps}. The number of elements of each attention map equals to the corresponding channel number of the feature matrix and the values of the elements represent the scale factors acting on the original features. Due to space limitation, only the 16-th to the 48-th channels are displayed here. A larger scale factor indicates more important channel of features. From the figure, we have the following observations:
\begin{itemize}
\item Without HAD, the role of attention is different in different depths of the attention-aided CNN. Specifically, as is shown in the first two subfigures, features are scaled in an angle-agnostic manner in shallower layers with small differences among average attention maps of different sine value ranges while the distributions of average attention maps become increasingly angle-specific in deeper layers. Notice that, the mean value of the 38-th scale factor of the third attention map varies significantly with sine value ranges. Reasonably, it can be inferred as a key angle-related feature in the considered problem. Such a phenomenon is also consistent with a typical discipline in DNNs that earlier layer features are more general while later layer features exhibit greater specificity\cite{feature_discipline}.
\item The distributions of average attention maps of closer sine value ranges are more similar. From the second subfigure, the curves of the first two ranges are very close to each other, while the curve of the third range is apparently different from them. It can be regarded as the embodiment of ``divide-and-conquer" since the channel estimation for data samples in the first two ranges and the third range can be regarded as two different subproblems, which are ``divided" by different attention maps first and then ``conquered" subsequently.
\item As is illustrated in the third subfigure, all scale factors in the fourth attention map are 0.5, which is due to the zero output of the former ReLU activation function and the Sigmoid activation function used to predict the attention map. Therefore, the last attention module is actually useless and can be removed during testing to further reduce the complexity\cite{attention}. 
\item From the fourth subfigure, the differences of average attention maps between sine value ranges are bigger and the binarization level of scale factors is higher in the HAD scenario. Only one attention module is used in the attention-aided FNN, so the ``divide" process has to be realized more intensely, which is different from the attention-aided CNN used in the scenario without HAD. Another reason might be that compared with the denoising process in the former scenario, reversing the effect of $\bm{W}$ is more angle-related, therefore the ``divide-and-conquer" policy is reflected more fully. When dealing with a certain subproblem, only specific features are kept and others are totally abandoned.
\end{itemize}

\begin{figure}[htb!] 
	\centering  
	\vspace{-0.35cm} 
	\subfigtopskip=2pt 
	\subfigbottomskip=1pt 
    \subfigcapskip=-3pt 
	\subfigure[The third attention map without HAD]{
		\label{single_att_map_3}
		\includegraphics[width=0.48\textwidth]{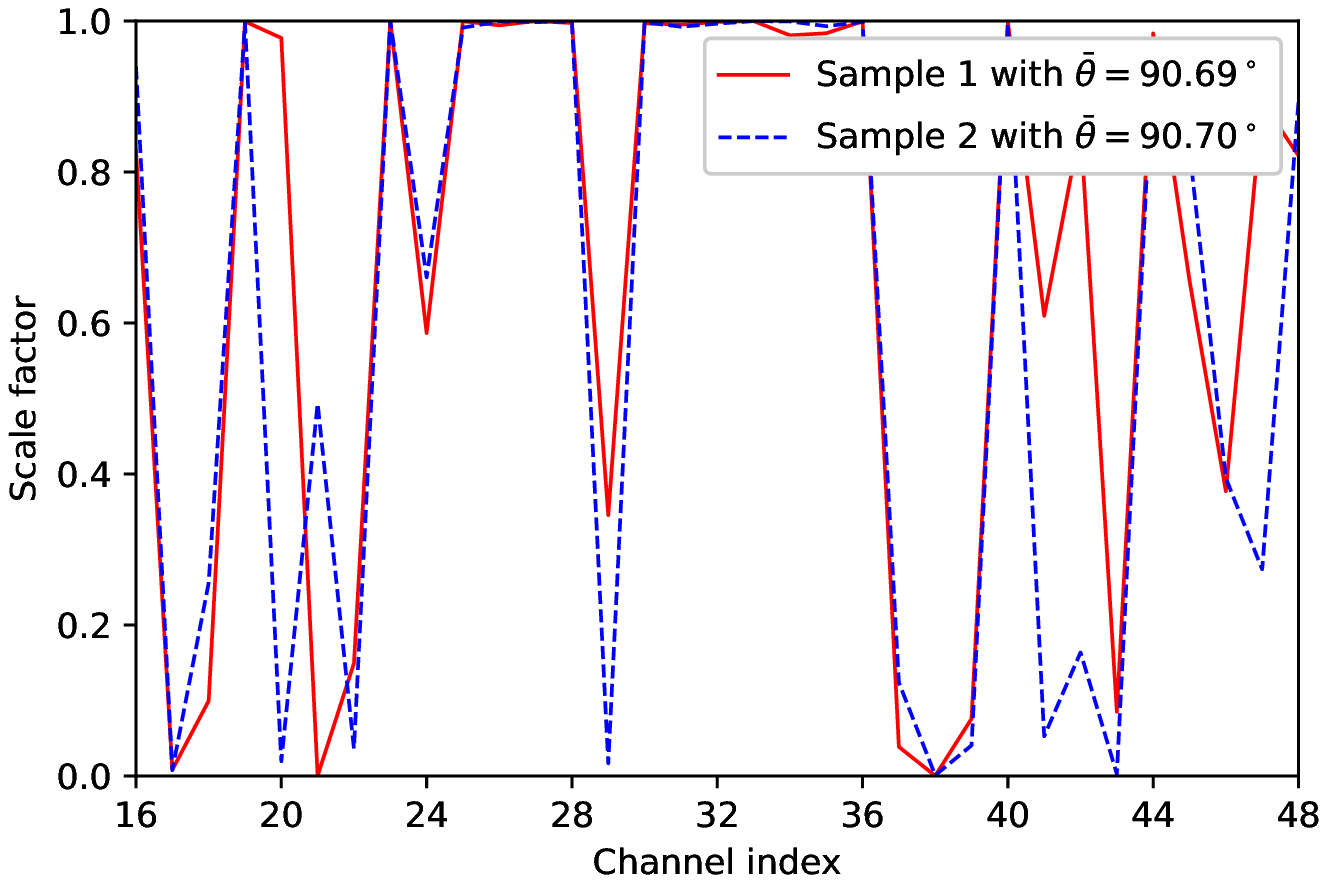}}
	\subfigure[The attention map with HAD]{
		\label{single_att_map_CS}
		\includegraphics[width=0.48\textwidth]{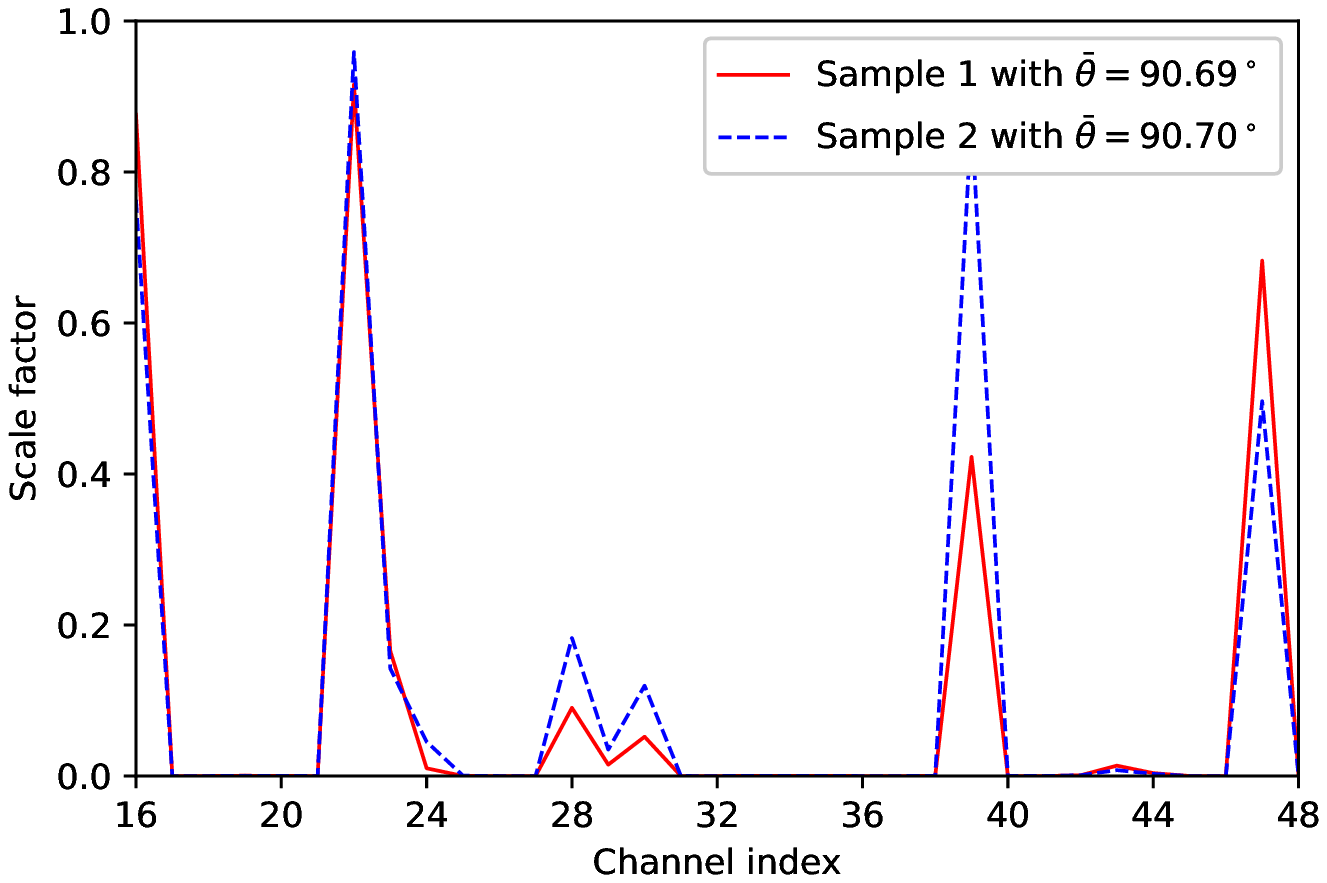}}
	\caption{The attention maps of two exemplary data samples with very close average AoAs.}
	\label{CE_sample_att}
\end{figure}

Apart from the statistical characteristics, Fig. \ref{CE_sample_att} also presents the attention maps of two exemplary data samples with close average AoAs. Although the average AoAs are almost same, the attention maps of these two data samples are still dramatically different, which reveals the sample-specific nature of attention. The reason is that although average AoA can reflect most of the channel's characteristics, there are still some features, such as the specific AoAs and gains of channel paths, which can also be exploited by attention for further performance improvement.

\subsection{Complexity Comparison}
Under typical system settings where $N=128$, $M=32$, $L_p=K=10$, and $I_E=50$, the specific complexity of different algorithms is compared in Table \ref{complexity}. Notice that the last attention layer in attention-aided CNN is removed during testing as mentioned above. Besides, for MMSE $3^\circ$, CCMs computed by channel samples whose average AoAs have same sine values can be shared to halve the number of parameters. 

As we can see, without HAD, the number of parameters only increases $19.86\%$ with the use of attention, and the additional FLOPs overhead introduced by attention is almost negligible. Although the FLOPs of attention-aided CNN are slightly higher than MMSE currently, it will be much smaller than MMSE if the antenna number keeps growing. Besides, the parameter number of MMSE $3^\circ$ is also quite large since tens of CCMs are required to exploit the narrow angular spread characteristic of channels.

In the scenario with HAD, we only compare three algorithms with practical performance. Both attention-aided CNN and FNN have similar parameter numbers while the FLOPs of attention-aided FNN is much lower. Remember that, its performance is also better than attention-aided CNN, which indicates the superiority of the proposed design. The FLOPs of S-VBI is significantly higher than the DL-based methods. In simulation, when both run on CPU, attention-aided FNN can be hundreds of times faster than S-VBI in terms of clock time and the advantage is even more exaggerated if accelerated by GPU.
\begin{table}[!htbp]
\centering
\begin{tabular}{|c|c|c|c|c|c|c|}
\hline
\multirow{2}{*}{Metrics} & \multicolumn{3}{c|}{Without HAD} & %
    \multicolumn{3}{c|}{With HAD}\\
\cline{2-7}
 & CNN & \bfseries{CNN+Att} & MMSE $3^\circ$ & CNN+Att & \bfseries{FNN+Att} & S-VBI \\
\hline
FLOPs ($\times10^7$) & 1.794 & \bm{1.801} & 1.689 & 0.531 & \bm{0.103} & 5.516\\
\hline
Parameters ($\times10^6$) & 0.141 & \bm{0.169} & 1.966 & 1.002 & \bm{1.072} &0 \\
\hline
\end{tabular}
\caption{Complexity comparison of algorithms under typical system settings.}
\label{complexity}
\end{table}

\section{Conclusion}
In this paper, we have proposed a novel attention-aided DL framework for massive MIMO channel estimation. Both the scenarios without and with HAD are considered and scenario-specific neural networks are customized correspondingly. By integrating the attention mechanism into CNN and FNN, the narrow angular spread characteristic of channel can be effectively exploited, which is realized by the ``divide-and-conquer" policy to dynamically adjust attention maps. The proposed approach can significantly improve the performance but is with relatively low complexity.

\nocite{*}
\bibliographystyle{IEEE}
\begin{footnotesize}

\end{footnotesize}

\end{document}